\documentclass[journal]{IEEEtran}

\usepackage{cite}
\usepackage[numbers,sort&compress]{natbib}
\usepackage{amsmath,amssymb,amsfonts}
\usepackage{graphicx}
\usepackage{subfigure}
\usepackage{textcomp}
\usepackage{xcolor}
\usepackage{bm}
\usepackage{caption}
\usepackage{algpseudocode}  

\usepackage{algorithm}

\usepackage{amsthm}

\newtheorem{proposition}{Proposition}

\newtheorem{remark}{Remark}
\usepackage{stfloats}

\def\BibTeX{{\rm B\kern-.05em{\sc i\kern-.025em b}\kern-.08em
    T\kern-.1667em\lower.7ex\hbox{E}\kern-.125emX}}

\ifCLASSINFOpdf
\else
\fi

\newcounter{TempEqCnt}

\begin{document}

\linespread {0.97}  
\addtolength{\parskip}{0.96pt} 

\setlength{\columnsep}{9pt}

\makeatletter
\renewcommand\normalsize{%
\@setfontsize\normalsize\@xpt\@xiipt
\abovedisplayskip 4\p@ \@plus2\p@ \@minus5\p@
\abovedisplayshortskip \z@ \@plus3\p@
\belowdisplayshortskip 6\p@ \@plus3\p@ \@minus3\p@
\belowdisplayskip \abovedisplayskip
\let\@listi\@listI}
\makeatother

\title{Synesthesia of Machines (SoM)-Enhanced \\ ISAC Precoding for Vehicular Networks \\ with Double Dynamics}


\author{Zonghui~Yang,~\IEEEmembership{Graduate~Student~Member,~IEEE},
Shijian~Gao,~\IEEEmembership{Member,~IEEE},\\
Xiang~Cheng,~\IEEEmembership{Fellow,~IEEE},
Liuqing~Yang,~\IEEEmembership{Fellow,~IEEE}
\thanks{Part of this paper has been accepted by the 2024 IEEE Global Communications Conference 
(GLOBECOM) \cite{GC}.}
\thanks{Z.~Yang and X.~Cheng are with the State Key Laboratory of Advanced Optical Communication Systems and Networks, School of Electronics, Peking University, Beijing 100871, China (e-mail: yzh22@stu.pku.edu.cn; xiangcheng@pku.edu.cn).}
\thanks{S.~Gao is with the Internet of Things Thrust, The Hong Kong University of Science and Technology (Guangzhou), Guangzhou 511400, China (e-mail: shijiangao@hkust-gz.edu.cn).}
\thanks{L.~Yang is with the Internet of Things Thrust \& Intelligent Transportation Thrust, The Hong Kong University of Science and Technology (Guangzhou) Guangzhou, China, and also with the Department of Electronic and Computer Engineering, Hong Kong University of Science and Technology, Hong Kong SAR, China (Email: lqyang@ust.hk).}
}
\maketitle

\markboth{IEEE TRANSACTIONS ON COMMUNICATIONS, 2024} %
{Shell \MakeLowercase{\textit{et al.}}: Bare Demo of IEEEtran.cls for IEEE Journals}

\begin{abstract}

Integrated sensing and communication (ISAC) technology is vital for vehicular networks, yet the time-varying communication channels and rapid movement of targets present significant challenges for real-time precoding design. Traditional optimization-based methods are computationally complex and depend on perfect prior information, which is often unavailable in double-dynamic scenarios. In this paper, we propose a synesthesia of machine (SoM)-enhanced precoding paradigm that leverages modalities such as positioning and channel information to adapt to these dynamics. Utilizing a deep reinforcement learning (DRL) framework, our approach pushes ISAC performance boundaries. We also introduce a parameter-shared actor-critic architecture to accelerate training in complex state and action spaces. Extensive experiments validate the superiority of our method over existing approaches.


\end{abstract}


\IEEEpeerreviewmaketitle

\vspace{-0.0cm}
\section{Introduction}

\IEEEPARstart{T}HE Internet of vehicles (IoV) plays a critical role in facilitating an efficient and reliable intelligent transportation system, where high-speed communication and high-precision sensing serve as fundamental pillars \cite{cheng2022integrated}. In contrast to conventional separated systems, integrated sensing and communication (ISAC) holds the promise of mutual enhancement, thus garnering significant attention in recent research endeavors within the realm of vehicular networks \cite{overview_JRC_for_AV}. 

To achieve ISAC, waveform design serves as one of the core techniques.
Existing dual-functional waveform designs are typically categorized into three types: communication-centric, radar-centric, and joint design. Communication-centric designs focus on enhancing sensing based on existing communication waveforms \cite{commu_cen_1,commu_cen_2}. 
However, the sensing capability is inherently limited due to the unregulated auto-correlation properties of communication symbols.
In radar-centric designs, information is embedded into radar waveform parameters such as carrier frequency and antenna assignment \cite{majorcom, FRaC}. 
Nevertheless, these designs are constrained by radar pulse repetition rate, failing to satisfy the high-rate requirements in practical communications.
To improve the sensing accuracy while guaranteeing communications, joint precoding designs by shaping the beam into desired patterns have been widely researched \cite{FanLiu_MUMIMO, FanLiu_towards, Joint_opt_ISAC}. For instance \cite{FanLiu_MUMIMO} achieved the desired pattern by optimizing the precoding, satisfying the signal-to-noise ratio (SNR) constraint at the user. \cite{Joint_opt_ISAC} achieved a higher sensing SNR while guaranteeing signal-to-interference-plus-noise ratio (SINR) at users by jointly optimizing the precoding for communication and radar signal.

However, the dynamic nature of the road environment introduces both time-varying characteristics in the wireless channel and rapid movement of the targets, presenting a doubly-dynamic challenge for ISAC.
To deal with the time-varying channels, \cite{dynamic_com_1} proposed a compressive sensing (CS)-based framework with specialized pilot, reducing the time overhead in channel state information (CSI) acquisition. \cite{ICI_wangjintao, fan2021wideband, radar_V2V} designed pre-compensators to combat the Doppler effect, mitigating the inter-carrier interference (ICI) in wideband systems.
\cite{Doppler_comp_angle_domain} further decoupled the Doppler shifts in angle domain by precoding to effectively suppress the time-variation of communications.
In terms of sensing the fast-moving targets,  \cite{CSI_based_Doppler_estimation} utilized  CSI to estimate the target's velocities, and \cite{dynamic_sensing_3} designed space-delay-doppler precoding by leveraging ICI to enhance the sensing accuracy.
Meanwhile, to address the carrier frequency offset resulting from the sensing dynamics, \cite{dynamic_sensing_4} designed a Kalman filter (KF)-based sensing scheme using the estimated Doppler as a prior to filter out time-varying noise and enhance sensing accuracy.

Nevertheless, implementing ISAC under double dynamics remains unexplored and presents unique challenges. 
Firstly, the intricate correlation and differences between the dynamics of communication and sensing complicate the simultaneous tracking of both users and targets, resulting in degraded spectrum efficiency (SE) and sensing accuracy.
Secondly, existing optimization-based methods are complex and result in high latency. Additionally, the requirement for perfect environmental information, such as instantaneous CSI and real-time target positions, can be challenging to fulfill in dynamic environments.

In an effort to lower the dependence on instantaneous state information and mitigate the complexity issue, \cite{dynamic_PF,dynamic_MP} predicted the real-time environmental prior according to the historical positions, via particle filter (PF) and message passing (MP) respectively. The operating protocol have also been modified for facilitating the predictive precoding. However, these solutions still struggle to address the doubly-dynamic challenge, as they lack robustness in managing the mutual interference between communication and sensing. Moreover, their complexity escalates drastically with a growing number of users, thereby reducing the timeliness of ISAC systems. Additionally, these solutions assume perfect alignment between communication users and sensing targets, which may not be a realistic assumption in real-world doubly-dynamic scenarios.

Although obtaining perfect channel-related prior in double dynamics is challenging, other observable environmental information correlates with channel information, as demonstrated in synesthesia of machine (SoM) \cite{overview_Som}. For instance, \cite{PL_SoM} predicted the real-time path loss in communication channels based on multi-modal non-CSI data, and \cite{MMFF_Som} integrated multi-modal sensing information to boost beamforming for communications. 
However, research from this perspective is still in its early stages, and specific SoM operational modes for ISAC remain undefined.
The key is to effectively utilize the observations from diverse modalities, such as position information and long-timescale CSI, which can be obtained with significantly lower overhead, to assist in real-time ISAC precoding design.
Unlike the location-assisted designs in \cite{location_aided_huang, location_aware_vtc}, the inherent sensing capability of the BS can be leveraged to achieve simultaneous communication and positioning, eliminating the need for additional sensors.

Reinforcement learning (RL) has gained attention in communication systems design thanks to its adaptability to dynamics without prior assumptions. Existing research has employed RL for precoding design \cite{ PrecoderNet, Fast_beamforming}. For instance, deep deterministic policy gradient (DDPG) was utilized for the hybrid precoding transceivers design in \cite{PrecoderNet}. However, these studies primarily focus on enhancing communications and overlook sensing tasks.
Regarding sensing applications, \cite{RL_radar_ISAC} employed SARSA algorithm to jointly optimize precoders and array segmentation for balancing communications and sensing. \cite{RL_time_divide} focused on flexible alternation between communication and sensing modes. Despite these efforts, the level of integration remains limited.
Furthermore, applying RL to the considered problem encounters three main challenges: the high-dimensional and structurally complex action space, the significant difficulties in feature extraction and action selection due to diverse environmental state observations, and the challenges in updating policies due to the varying number of users.

To overcome the aforementioned barriers, we devise an efficient deep-RL (DRL)-aided framework to facilitate real-time wideband ISAC precoding design, in the absence of instantaneous CSI or target angle information. 
Inspired by SoM, this framework leverages various observations from the base station (BS), such as historical position estimates and initial CSI, as adaptable inputs to accommodate the double dynamics.
We begin by formulating a partially observed Markov decision process (POMDP) for this scenario, defining the observed state space and reward function. To navigate the intricate decision space in hybrid precoding, we customize a hybrid action space for efficient precoding updates. Furthermore, we employ a parameter-shared actor-critic (PSAC) architecture to extract features from the state, reducing network parameters and speeding up training. To optimize model training, we tailor the updating procedure for both the actor and critic components. Our proposed algorithm incorporates two carefully crafted precoding update schemes: update across the user dimension and update across the antenna dimension.
It is noteworthy that the proposed scheme also exemplifies SoM-Enhance in \cite{overview_Som}, wherein channel information for both communication and sensing is enriched through observations from various modalities.
Numerical experiments validate the effectiveness of our approach in improving both communications and sensing performance in a doubly dynamic environment.

Our contributions can be summarized as follows:
\begin{itemize}
    \item We present an alternating optimization-based algorithm for ISAC hybrid precoding with complete environmental state information, designed to adapt to ICI caused by Doppler shift in wideband multi-user systems.
    \item To tackle the doubly-dynamic ISAC precoding challenge, we introduce a DRL-aided algorithm. We carefully craft the network architecture and training process to address the complex decision space issue. The dynamic update schemes proposed are compatible with existing hybrid precoding architectures.
    \item Through extensive simulation experiments, we demonstrate that our approach outperforms existing benchmarks, pushing the performance boundaries for ISAC in doubly-dynamic scenarios. This is achieved with reduced computational complexity and improved timeliness.
\end{itemize}  

\newcommand{\RNum}[1]{\uppercase\expandafter{\romannumeral #1\relax}}

The remainder of this work is structured as follows. Section \RNum{2} introduces the system model. The proposed optimization-based wideband ISAC precoding design is illustrated in Section \RNum{3} as preliminary. Section \RNum{4} includes the ISAC frame structure designed for doubly-dynamic scenarios, and the detailed DRL-aided algorithms for doubly-dynamic ISAC precoding is introduced in Section \RNum{5}. Section \RNum{6} contains our simulation findings, and Section \RNum{7} concludes our work.

\textit{Notation}: $a$, $\bm a$ and $\bm A$ represent a scalar, a vector and a matrix respectively. $(\cdot)^{\text{T}}$, $(\cdot)^{\text{H}}$, $\text{Tr}(\cdot)$, $\text{rank}(\cdot)$, $\left\|\cdot\right\|_2$ and $\left\|\cdot\right\|_{\text{\text{F}}}$ denote transpose, conjugate transpose, trace, rank, 2-norm and Frobenius norm, respectively. $\mathbb{E}(\cdot)$ denotes the expectation. $\mathbb{P}(\cdot)$ denotes the probability. $\mathcal{CN}(m,\sigma^2)$ represents the complex Gaussian distribution whose mean is $m$ and covariance is $\sigma^2$. $\mathbb{R}$ and $\mathbb{C}$ denote the set of real numbers and complex numbers respectively. $\vert z \vert$ and $\angle z$ denote the modulus and the phase of a complex number $z$ respectively. $\bigcap$ and $\bigcup$ denote the intersection and the union, and $\emptyset$ denotes an empty set.

\vspace{-0.1cm}
\section{System Model}
We consider an ISAC system with wideband massive MIMO at the base station (BS) serving $U$ downlink users, each with a single antenna, while simultaneously tracking a single moving target in the low-altitude environment. The BS is equipped with a uniform linear array (ULA) comprising $N_{\text{t}}$ elements for transmission and $N_{\text{r}}$ elements for echo reception. The ISAC signal is transmitted via orthogonal frequency division multiplexing (OFDM) with $M$ subcarriers.
The number of data streams at each subcarrier is $N_{\text{s}}$, satisfying $U\leqslant N_{\text{s}}$ and $N_{\text{s}}\leqslant N_{\text{RF}}$, $(m=1,\cdots, M)$. The sensing and communication functionalities are operated simultaneously enabled by an ISAC transmitter.
Without loss of generality, we set $N_{\text{s}}=N_{\text{RF}}=U$.
One ISAC frame is divided into $T$ subframes, each containing $L$ OFDM symbols. We assume that the users and target' locations remain invariable within one subframe, and their velocities are constant within one frame.

\begin{figure}[t]
  \vspace{-0.0cm}
  \centering
  \includegraphics[width=0.98\linewidth]{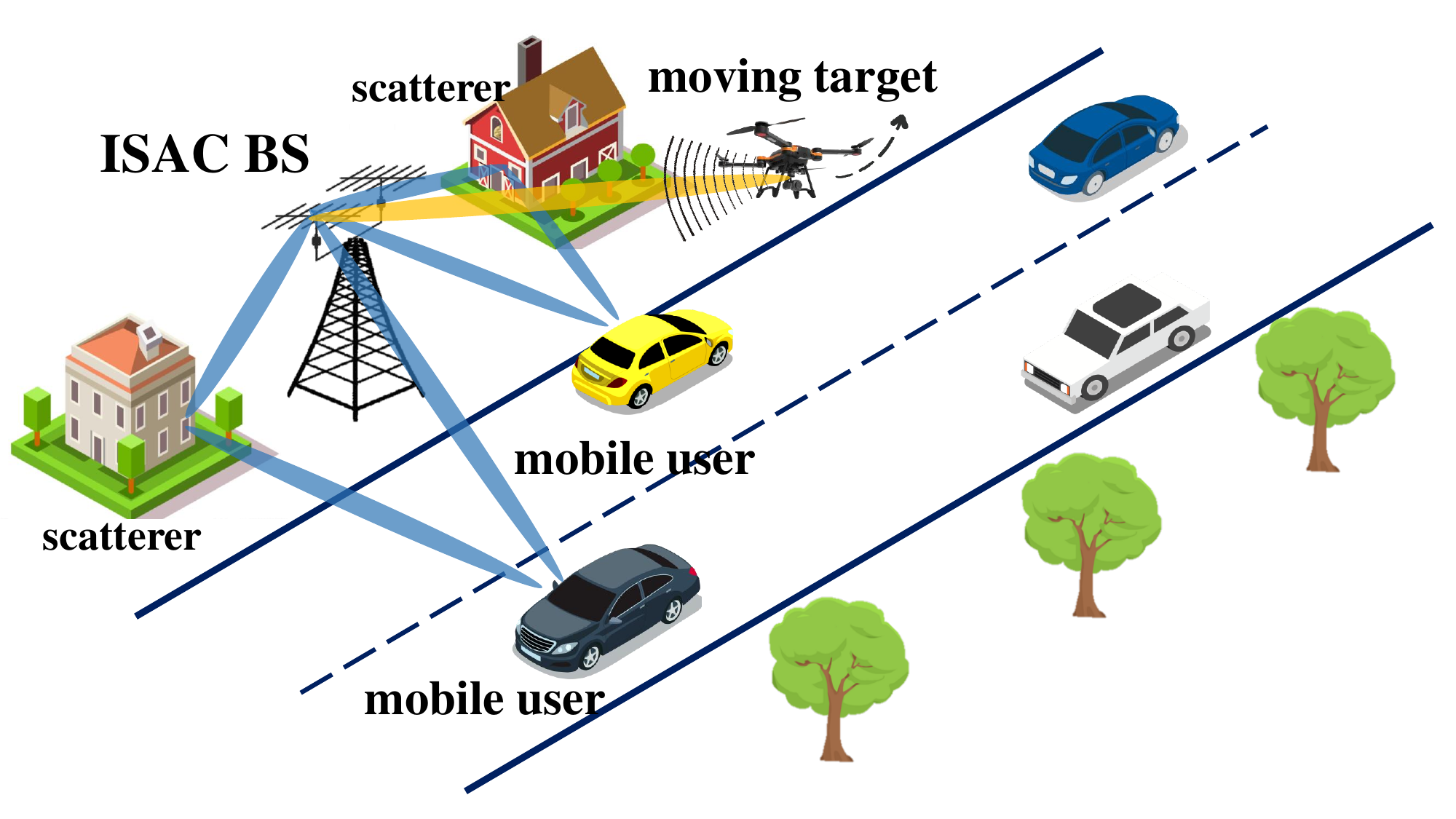}
  \vspace{-0.1cm}
  \captionsetup{font={small}}
  \caption{Investigated system model where an ISAC BS serves multiple vehicles and senses a moving target.}
  \vspace{-0.3cm}
  \label{fig:com model}
\end{figure}

\vspace{-0.2cm}
\subsection{Transmitting Signal}

At the $l$-th symbol duration in the $n$-th subframe ($l=1,\cdots, L$, $n=1,\cdots, T$), the communication symbols for the $U$ users at the $m$-th subcarrier is denoted as $\bm s_{m}^{(n,l)}=\left[ s_{m,1}^{(n,l)},\cdots, s_{m,U}^{(n,l)}\right ]^{\text{T}}$ with $\mathbb{E} \left [\bm s_{m}\bm s_{m}^{\text{H}}\right ]=\bm I_{\text{U}}$.
Specifically, $\bm s=\left[ \bm s_{1}^{\text{T}},\cdots, \bm s_{\text{M}}^{\text{T}} \right]^{\text{T}}$ is processed by $M$ frequency-dependent digital precoders $\bm F_{\text{BB},m}\in \mathbb{C}^{N_{\text{RF}}\times U}$ $(m=1,2,\cdots,M)$ respectively, followed by $N_{\text{RF}}$ $M$-point inverse fast Fourier transforms (IFFT). The the $L_{\text{CP}}$-length cyclic prefix (CP) is then added on the time-domain signal, before the time-domain analog precoder $\bm F_{\text{RF}}\in \mathbb{C}^{N_{t}\times N_{\text{RF}}}$ which is shared among all the subcarriers. 
Thus the discrete-time transmitted signal at the $m$-th subcarrier from the array in the $n$-th subframe can be expressed as
\vspace{-0.2cm}
\begin{equation}
    \bm x_{m}^{(n,l)}=\sqrt{P_{\text{t}}}\bm F_{\text{RF}}^{(n)}\bm F_{\text{BB},m}^{(n)}\bm s_{m}^{(n,l)}=\sqrt{P_{\text{t}}}\bm F_{m}^{(n)}\bm s_{m}^{(n,l)},
    \vspace{-0.2cm}
\end{equation}
where $P_{\text{t}}$ denotes the total transmitting power. The digital precoder $\bm F_{\text{BB},m}$ is set as $\text{diag}(\left[ p_{m,1}, \cdots, p_{m,U} \right])$, used for power allocation among users.
Since the analog precoder $\bm F_{\text{RF}}$ is realized by the phase shifters (PS), it is constant-modulus. Specifically, when the $B$-bit PS is adopted, we have 
\vspace{-0.1cm}
\begin{equation}
    \angle (\bm F_{\text{RF}}[i,j])\in \mathcal{B}\triangleq \left\{ \frac{2\pi b}{2^{B}} \vert b=0,1,2,\cdots,2^{B}-1 \right\}.
\vspace{-0.2cm}
\end{equation}
$\mathcal{F}$ denotes the feasible set for all the $\bm F_{\text{RF}}$ satisfying the $B$-bit finite-resolution constant-modulus constraint.

\subsection{Communication Model}
A doubly-selective channel similar to \cite{DSDS_TWC} is adopted, with $N_{\text{d}}$ delay taps. At the $d$-th tap $(0\leqslant d \leqslant N_{\text{d}}-1)$, the channel response for user-$u$ during symbol-$l$ is expressed as
\vspace{-0.2cm}
\begin{equation}
\widetilde{\bm h}_{u,d}(l)=\sqrt{\frac{N_{\text{t}}}{P_{\text{u}}}}\sum_{p=1}^{P_{\text{u}}}\beta_{u,p}g_{\text{rp}}(dT_{\text{s}}-\tau_{u,p})e^{j\omega_{u,p}l} \bm a_{\text{t}}(\theta_{u,p}),
\vspace{-0.2cm}
\end{equation}
where $P_{\text{u}}$ denotes the number of resolvable paths, $T_{\text{s}}$ is the sampling period, $\beta_{u,p}\sim\mathcal{CN}(0,\sigma_{\beta}^{2})$ is the path's complex gain varying across symbols and $\theta_{u,p}$ is the angle of departure (AoD) of the $p$-th path for the $u$-th user, and $\bm a_{\text{t}}(\theta)[i]=\frac{1}{\sqrt{N_{\text{t}}}}e^{-j(i-1)\pi \sin{\theta}}$ $(i=1,\cdots,N_{\text{t}})$.
$g_{\text{rp}(\cdot)}$ is the pulse shaping filter; $\tau_{u,p}\sim \mathcal{U}(0,(N_\text{d}-1)T_{\text{s}})$ is the propagation delay. Denote the carrier frequency as $f_{\text{c}}$, the velocity of light as $c_{\text{v}}$, and the relative velocity of each path as $\bm v_{u,p}$, then the normalized Doppler shift is $\omega_{u,p}=2\pi f_{\text{c}}\vert \bm v_{u,p}\vert T_{\text{s}}\sin{\theta_{u,p}}/c_{\text{v}}$.
Consequently the multi-user channel in time domain is 
\vspace{-0.1cm}
\begin{equation}
    \widetilde{\bm H}_{d}(l)=\left [ \widetilde{\bm h}_{1,d}^{\text{T}}(l),\widetilde{\bm h}_{2,d}^{\text{T}}(l),\cdots,\widetilde{\bm h}_{U,d}^{\text{T}}(l)  \right ]^{\text{T}}\in \mathbb{C}^{U\times N_{\text{t}}}.
    \vspace{-0.2cm}
\end{equation}
Accordingly, the inter-subcarrier channel response is computed by
\vspace{-0.1cm}
\begin{equation}
    \bm H_{m}[k]=\frac{1}{M}\sum_{i=0}^{M-1}\sum_{d=0}^{N_{\text{d}}-1}\widetilde{\bm H}_{d}(i+L_{\text{CP}})e^{-j2\pi\frac{kd+(m-k)i}{M}}.
    \vspace{-0.2cm}
\end{equation}
By stacking the received signals at all subcarriers, we get
\vspace{-0.1cm}
\begin{align}
&\bm y_{\text{c}}=\left[  \bm y_{1}^{\text{T}}, \bm y_{2}^{\text{T}}, \cdots, \bm y_{M}^{\text{T}} \right]^{\text{T}} \\ \nonumber
&=
\!\sqrt{P_{\text{t}}}\!\underbrace{
\left[
\begin{matrix}
\bm H_{1}[1] \! &\! \cdots\! & \!\bm H_{1}[M]\\
\vdots \!& \!\ddots \!&\! \vdots \\
\bm H_{M}[1] \!&\! \cdots \!& \!\bm H_{M}[M]
\end{matrix}
\right]
}_{\overline{\bm H}}
\underbrace{
\left[ 
\begin{matrix}
\bm F_{1} \!&\! \! &\!    \\
 &\!   \ddots \!&\!  \\
 &\! \!&\! \bm F_{M} \\
\end{matrix}
\right]
}_{\overline{\bm F}}
\left[
\begin{matrix}
\bm s_{1}\\ \vdots \\ \bm s_{M}
\end{matrix}
\right]\!\!+\!\!\bm n_{\text{c}},
\end{align}
where $\bm n_{\text{c}}\sim \mathcal{CN}(\bm 0,\sigma_{\text{c}}^{2}\bm I_{UM})$ is the additive white Gaussian noise (AWGN). 

\begin{table}[t]
    \centering
    \caption{Notations related to the ISAC system.}
    \begin{tabular}{|p{2.65cm}||p{5.45cm}|}
    \hline
       Notation  &  Definition \\
       \hline
       \hline
       $\widetilde{\bm h}_{u,d}(l)\in \mathbb{C}^{1\times N_{\text{t}}}$  &  Communication channel per user in time domain \\
       \hline
       $\widetilde{\bm H}_{d}(l)\in \mathbb{C}^{U\times N_{\text{t}}}$  &  Overall communication channel in time domain \\
       \hline
       $\bm H_{m}[k]\in \mathbb{C}^{U\times N_{\text{t}}}$  &  Inter-carrier response in frequency domain  \\
       \hline
       $\overline{\bm H}\in\mathbb{C}^{UM\times N_{\text{t}}M}$  &  Wideband channel in frequency domain \\
       \hline
       $\bm G_{m}\in\mathbb{C}^{N_{\text{t}}\times N_{\text{t}}}$  &  Target response at the $m$-th subcarrier \\
       \hline
       $\bm F_{\text{BB},m}\in \mathbb{C}^{U\times U}$  &  Frequency-dependent digital precoding matrix \\
       \hline
       $\bm F_{\text{RF}}\in \mathbb{C}^{N_{\text{t}}\times U}$  &  Frequency-independent analog precoding matrix \\
       \hline
       $\bm F_{m}\in \mathbb{C}^{N_{\text{t}}\times U}$  &  Equivalent digital precoding matrix \\
       \hline
       $\overline{\bm F}\in \mathbb{C}^{N_{\text{t}}M\times UM}$  &  Wideband overall precoding matrix \\
       \hline
       $\bm h_{m,k,u}\!=\!\bm H_{m}[k][u,:]$  &   Frequency-domain inter-carrier channel per user \\
       \hline
       $\bm f_{m,u}=\bm F_{m}[:,u]$  &  Equivalent precoding per user per subcarrier \\
       \hline
    \end{tabular}
    \label{tab:notation}
    \vspace{-0.3cm}
\end{table}

Note that at the $m$-th subcarrier, the $u$-th user receives not only the customized communication signals but also the inter-user interference (IUI) and inter-carrier interference (ICI). Hence, the corresponding signal-to-interference-plus-noise ratio (SINR), $\gamma_{\text{u,m}}^{(n,l)}$, can be derived as
\vspace{-0.1cm}
\begin{equation}
\!\frac{|\bm h_{m,m,u}^{(n,l)}\bm f_{m,u}^{(n)}|^{2}}{\sum_{k\neq m}\!\sum_{u=1}^{U}\!|\bm h_{m,k,u}^{(n,l)}\bm f_{k,u}^{(n)}\!|^{2}\!+\!\sum_{i \neq u}\!|\bm h_{m,m,u}^{(n,l)}\bm f_{m,i}^{(n)} \!|^{2} \!+\!\frac{N_{\text{t}}UM\sigma_{\text{c}}^{2}}{P_{\text{t}}}},
\end{equation}
where $\bm f_{m,u}=\bm F_{m}[:, u]$ and $\bm h_{m,k,u}=\bm H_{m}[k][u,:]$. Consequently, the spectrum efficiency (SE) can be obtained as
\vspace{-0.1cm}
\begin{equation}
 \mathcal{R}^{(n)}=\frac{\sum_{l=1}^{L}\sum_{m=1}^{M}\sum_{u=1}^{U}\log_{2}(1+\gamma_{\text{u,m}}^{(n,l)})}{LT_{\text{s}}M \Delta f},
 \label{equ:SE}
\end{equation}
which is a crucial metric of the quality of services (QoS) for communications. $\overline{\mathcal{R}}=\frac{1}{T}\sum_{n=1}^{T}\mathcal{R}^{(n)}$ is the averaged SE to be optimized during the whole frame.

\vspace{-0.2cm}
\subsection{Sensing Model}
In the ISAC system, the transmitted signal can serve for target sensing. The variations in the echo carry valuable information about the targets. A dedicated echo processing module can be incorporated to extract target information at the transmitter, akin to a mono-static MIMO radar system.

We model the target's trajectory as a unidirectional curve with an orientation offset.
$\left[ \theta_{\text{s}}^{(n)},d_{\text{s}}^{(n)} \right ]$ denotes the target's position in polar domain.
The target response matrix at the $m$-th subcarrier $\bm G^{(n)}_{m}=\alpha^{(n)}_{m}\bm a_{\text{r}}(\theta_{\text{s}}^{(n)})\bm a_{\text{t}}^{\text{H}}(\theta_{\text{s}}^{(n)})$, where $\theta_{\text{s}}^{(n)}$ is the azimuth of the target relative to the BS and $\alpha_{m}^{(n)}$ denotes the complex reflection coefficient incorporating the round-trip path loss, the radar cross-section and the Doppler of the target. 
Accordingly the received echo becomes
\vspace{-0.1cm}
\begin{equation}
    \bm y_{\text{r},m}^{(n,l)}=\bm G_{m}^{(n)} \bm x_{m}^{(n,l)}+\bm z_{m}^{(n,l)},
    \vspace{-0.2cm}
\end{equation}
with $\bm z_{m}^{(n,l)}\sim\mathcal{CN}(\bm 0, \sigma_{\text{s}}^{2}\bm I_{N_{\text{r}}})$ being the ISAC receiver noise. Note that BS also receives clutter reflected by objects unrelated to the target in the environment. We consider whitening the received noise to ensure that the noise obeys the complex Gaussian distribution.

\setcounter{TempEqCnt}{\value{equation}} 
\setcounter{equation}{9} 
\begin{figure*}[ht]
\vspace{-0.0cm}
\begin{equation}
\label{equ:CRB}
     \mathcal{J}_{m}(\theta_{\text{s}}^{(n)})=\frac{2\| \alpha_{m} \|^{2}(\text{Tr}(\dot{\bm A} ^{\text{H}}(\theta_{\text{s}}^{(n)})\dot{\bm A}(\theta_{\text{s}}^{(n)})\bm R_{x,m}^{(n)})\text{Tr}(\bm A^{\text{H}}(\theta_{\text{s}}^{(n)})\bm A(\theta_{\text{s}}^{(n)})\bm R_{x,m}^{(n)})-\|\text{Tr}(\dot{\bm A}^{\text{H}}(\theta_{\text{s}}^{(n)}) \bm A(\theta_{\text{s}}^{(n)}) \bm R_{x,m}^{(n)} )\|^{2})} {\sigma_{\text{s}}^{2}\text{Tr}(\bm A^{\text{H}}(\theta_{\text{s}}^{(n)})\bm A(\theta_{\text{s}}^{(n)})\bm R_{x,m}^{(n)})} .
\end{equation}
\hrulefill
\vspace{-0.2cm}
\end{figure*}

The sensing accuracy is characterized by Fisher information in Eq.~(\ref{equ:CRB}), with $\bm R_{x,m}^{(n)}=\frac{1}{L}\mathbb{E}\left[ \sum_{l=1}^{L} \bm x_{m}^{(n,l)}\bm x_{m}^{(n,l),\text{H}} \right ] =\frac{P_{\text{t}}}{N_{\text{t}}UM}\bm F_{\text{RF}}^{(n)}\bm F_{\text{BB},m}^{(n)}{\bm F_{\text{BB},m}^{(n)}}^{\text{H}}{\bm F_{\text{RF}}^{(n)}}^{\text{H}}$ being the covariance matrix of $\bm x_{m}$, and $\bm A(\theta_{\text{s}})=\bm a_{\text{r}}(\theta_{\text{s}})\bm a_{\text{t}}^{\text{H}}(\theta_{\text{s}})$ with $ \dot{\bm A}(\theta_{\text{s}})=\frac{\partial \bm A(\theta_{\text{s}})}{\partial \theta_{\text{s}}}$. The Fisher information for $\theta_{\text{s}}$ at $t$ is the summation of all the Fisher information across subcarriers, as analyzed in \cite{CRB_TAP}. Consequently, the Cramér-Rao lower bound (CRLB) of the target's azimuth is the expectation of the inverse of Fisher information given $\bm y_{\text{r}}$.
\vspace{-0.2cm}
\begin{equation}
    \text{CRLB}(\theta_{\text{s}}^{(n)})\!=\!\mathbb{E}\left [\frac{1}{\mathcal{J}(\theta_{\text{s}}^{(n)})}\right ]\!=\!\mathbb{E}\left [\frac{1}{\sum_{m=1}^{M}\mathcal{J}_{m}(\theta_{\text{s}}^{(n)})}\right ].
    \vspace{-0.2cm}
\end{equation}
\vspace{0cm}
Accordingly, the averaged CRLB over time, i.e., $\overline{\text{CRLB}}=\frac{1}{T}\sum_{n=1}^{T}\text{CRLB}(\theta_{\text{s}}^{(n)})$, is established as the performance metric for sensing.

\vspace{-0.0cm}
\section{Optimization-based ISAC precoding with instantaneous state}

In this section, we propose an optimization-based algorithm for hybrid precoding to implement ISAC in wideband multi-user system, assuming that full state information for sensing and communications can be obtained at the BS.

\vspace{-0.2cm}
\subsection{Related Works}

Extensive research has been conducted on precoding optimization for rate maximization \cite{shijian_MI} or CRLB optimization \cite{CRB_opt_radar_1}. However, ISAC precoding presents a greater challenge due to its non-convex constraints. Recent studies, such as \cite{ISAC_CRB_opt}, have focused on digital precoding optimization for CRLB minimization while ensuring communication quality of service (QoS) through the use of the semi-definite relaxation (SDR) method. Furthermore, \cite{ISAC_CRB_opt_subspace} delved deeper into CRLB optimization under SINR constraints from a subspace perspective, providing additional insights for digital precoding design. Addressing the non-convex constraint introduced by the hybrid structure, \cite{ISAC_hybrid_icc2023} conducted the data rate maximization under CRLB constraints by the block coordinate descent (BCD) algorithm. However, these studies do not take into account wideband time-varying channels, which is challenging to address due to the significant ICI resulting from Doppler.

\vspace{-0.1cm}
\subsection{ Alternating-based Optimization}
The goal of the hybrid precoding in this doubly-dynamic scenario is to maximize ISAC utility throughout the trajectory, considering the power budget and constant-modulus constraint of the analog precoder. The utility is defined as the weighted sum of the averaged spectrum efficiency $\overline{\mathcal{R}}$ and the averaged Fisher information for angle estimation $\overline{\mathcal{J}}=\frac{1}{T}\sum_{n=1}^{T}\sum_{m=1}^{M}\mathcal{J}_{m}(\theta_{\text{s}}^{n})$. Both indicators are normalized by the ideal performance boundaries $\overline{\mathcal{R}}^{*}=\frac{1}{T}\sum_{n=1}^{T}{\mathcal{R}^{(n)}}^{*}$ and $\overline{\mathcal{J}}^{*}=\frac{1}{T}\sum_{n=1}^{T}{\mathcal{J}^{(n)}}^{*}$ respectively, with ${\mathcal{R}^{(n)}}^{*}$ and ${\mathcal{J}^{(n)}}^{*}$ being the performance boundaries in the $t$-th subframe, obtained by individual maximization of SE without sensing constraints, and individual minimization of CRLB without communication requirements, respectively.
Consequently, the problem can be formulated as
\vspace{-0.2cm}
\begin{align}
    \underset{ \{\bm F_{\text{RF}}^{(n)}, \bm F_{\text{BB},m}^{(n)}\}_{n=1}^{T},  }{\text{max}}&~\psi \frac{\overline{\mathcal{R}}}{\overline{\mathcal{R}}^{*}} + (1-\psi) \frac{\overline{\mathcal{J}}}{\overline{\mathcal{J}}^{*}} \label{Problem0}\\
    \text{s.t.} &~ \sum_{m=1}^{M}\|\bm F_{\text{RF}}^{(n)}\bm F_{\text{BB},m}^{(n)}\|_{\text{F}}^{2}\leqslant 1, \tag{\ref{Problem0}{a}} \label{Problem0_a}\\
    &~ \bm F_{\text{RF}}^{(n)} \in \mathcal{F}, \tag{\ref{Problem0}{b}} \label{Problem0_b}
    \vspace{-0.3cm}
\end{align}
given the perfect $\overline{\bm H}$ and perfect $\bm G_{m}$ in each subframe. $\psi$ denotes the weighting coefficient between the sensing and communication performances. (\ref{Problem0_a}) and (\ref{Problem0_b}) denote the power constraint and the constant-modulus constraint respectively.
The problem (\ref{Problem0}) is non-convex due to the complex objective function and constant-modulus constraints.
Note that the time-averaged objective can be maximized when the utility at each single subframe is maximized.
We first transform the sequential optimization on a subframe-by-subframe basis as
\vspace{-0.2cm}
\begin{align}
    \underset{\bm F_{\text{BB},m}^{(n)}, \bm F_{\text{RF}}^{(n)}}{\text{max}} &~\psi \frac{\mathcal{R}^{(n)} }{ {\mathcal{R}^{(n)}}^{*}} +(1-\psi)\frac{\mathcal{J}^{(n)}}{{\mathcal{J}^{(n)}}^{*}} \label{Problem1} \\
    \text{s.t.} &~ \sum_{m=1}^{M}\|\bm F_{\text{RF}}^{(n)}\bm F_{\text{BB},m}^{(n)}\|_{\text{F}}^{2}\leqslant 1, \tag{\ref{Problem1}{a}} \label{Problem1_a}\\
    &~ \vert \bm F_{\text{RF}}^{(n)}[i,j]\vert =1. \tag{\ref{Problem1}{b}} \label{Problem1_b}
\end{align}
To tackle the coupled variables and non-convex constraints, we adopt an alternating-based optimization procedure. We first optimize the equivalent fully-digital precoding matrices $\bm F_{m}^{(n)}=\bm F_{\text{RF}}^{(n)}\bm F_{\text{BB},m}^{(n)}$ for ISAC utility maximization, then optimize $\bm F_{\text{RF}}^{(n)}$ and $\bm F_{\text{BB},m}^{(n)}$ by decomposing $\bm F_{m}^{(n)}$ in an alternating manner. Specifically, in each alternating iteration, the phases in $\bm F_{\text{RF}}^{(n)}$ are updated with fixed $\bm F_{\text{BB},m}^{(n)}$, followed by updating $\bm F_{\text{BB},m}^{(n)}$ with $\bm F_{\text{RF}}^{(n)}$ held constant.

\setlength{\parindent}{0pt}
\textbf{Stage \RNum{1}: Optimizing $\bm F_{m}$}
\setlength{\parindent}{10pt}

The problem with respect to $\bm F_{m}$ can be transformed as
\vspace{-0.1cm}
\begin{align}
    \underset{\bm F_{m}}{\text{max}} &~\psi \frac{\mathcal{R}}{ \mathcal{R}^{*}} +(1-\psi)\frac{\mathcal{J}}{\mathcal{J}^{*}} \label{Problem2} \\
    \text{s.t.} &~ \sum_{m=1}^{M}\|\bm F_{m}\|_{\text{F}}^{2}\leqslant 1.  \tag{\ref{Problem2}{a}} \label{Problem2_a}
    \vspace{-0.2cm}
\end{align}
Denote $\bm W_{m,u}=\bm f_{m,u}\bm f_{m,u}^{\text{H}}$, $\bm W_{m}=\sum_{u=1}^{U}\bm W_{m,u}$ and $\bm Q_{m,k,u}=\bm h_{m,k,u}^{\text{H}}\bm h_{m,k,u}$.
Then the problem (\ref{Problem2}) can be transformed into
\vspace{-0.3cm}
\begin{align}
    \underset{ \bm W_{m,u},\zeta,\tau  }{\text{max}}&~ \omega\label{Problem4}\\
    \text{s.t.} &~ \bm W_{m,u}\succeq \bm 0,~~\text{rank}(\bm W_{m,u})=1,~ \tag{\ref{Problem4}{a}} \label{Problem4_a} \\
    &~ \omega\leqslant \psi \frac{UM\log (1+\tau)}{\mathcal{R}^{*}}+(1-\psi) \frac{M\zeta} { \mathcal{J}^{*}}, \tag{\ref{Problem4}{b}} \label{Problem4_b}\\
    &~ \left [  \begin{matrix}  \text{Tr}(\Dot{\bm A}^{\text{H}}\Dot{\bm A}\bm W_{m})-\zeta  &  \text{Tr}(\Dot{\bm A}^{\text{H}}\bm A\bm W_{m}) \\  \text{Tr}(\bm A^{\text{H}}\Dot{\bm A}\bm W_{m}) &  \text{Tr}(\bm A^{\text{H}}\bm A\bm W_{m})  \end{matrix}   \right] \geq \bm 0, \tag{\ref{Problem4}{c}} \label{Problem4_c}\\
    &~ \text{Tr}(\bm Q_{m,m,u}\bm W_{m,u})-\tau \sum_{i=1,i\neq u}^{U}\!\text{Tr}(\bm Q_{m,m,u}\bm W_{m,i}) \nonumber \\
    &~-\tau \sum_{k=1, k\neq m}\bm Q_{m,k,u}\bm W_{k,u} \geqslant \tau \sigma_{\text{c}}^{2}, \tag{\ref{Problem4}{d}} \label{Problem4_d}\\
    &~ \sum_{m=1}^{M}\text{Tr}(\bm W_{m})\leqslant 1. \tag{\ref{Problem4}{e}} \label{Problem4_e}
\end{align}
where (\ref{Problem4_b}) represents for the SINR constraint equivalent to $\gamma_{m,u}\geqslant \tau$, since $\vert \bm h_{m,m,u}\bm f_{m,u}\vert^{2}=\text{Tr}(\bm h_{m,m,u}\bm W_{m,u}\bm h_{m,m,u}^{\text{H}})$. (\ref{Problem4_c}) represents for the sensing constraint equivalent to $\mathcal{J}_{m}(\theta_{\text{s}})\geqslant \zeta$, by applying Schur complement theorem. $\omega$ serves as the lower bound for the ISAC utility per subframe.
Subsequently, we employ the classical SDR technique by eliminating the rank-1 constraints in (\ref{Problem4_a}). This problem is then transformed into a standard semi-definite programming (SDP) and can be solved by the off-the-shelf toolbox like CVX.
As a result, the rank-1 constraint is restored as elaborated in \cite{Joint_opt_ISAC} and the equivalent fully-digital precoding matrix $\bm F_{m}$ can be obtained.

\vspace{-0.3cm}

\begin{proposition}
There exists a global optimum for the relaxed problem satisfying the rank-1 constraint, which ensures the tightness of the SDR can be guaranteed.
\end{proposition}
\vspace{-0.2cm}
\begin{proof}
    See Appendix A.
\end{proof}

\setlength{\parindent}{0pt}
\textbf{Stage \RNum{2}: Decomposing $\bm F_{\text{m}}$}
\setlength{\parindent}{10pt}

Upon obtaining $\bm F_{m}$, the remaining task is to optimize $\bm F_{\text{BB},m}$ and $\bm F_{\text{RF}}$ to approximate $\bm F_{m}$ as
\vspace{-0.2cm}
\begin{align}
    \underset{ \bm F_{\text{BB},m}, \bm F_{\text{RF}} }{\text{min}}&~\sum_{m=1}^{M}\|\bm F_{m}-\bm F_{\text{RF}}\bm F_{\text{BB},m} \|_{\text{F}}^{2} \label{Problem6}\\
    \text{s.t.} &~ \sum_{m=1}^{M}\|\bm F_{\text{RF}}\bm F_{\text{BB},m}\|_{\text{F}}^{2}=1, \tag{\ref{Problem6}{a}} \label{Problem6_a}\\
    &~ \bm F_{\text{RF}}\in \mathcal{F}. \tag{\ref{Problem6}{b}} \label{Problem6_b}
\end{align}

\vspace{-0.2cm}
\subsubsection{Optimizing $\bm F_{\text{RF}}$}

With fixed $\bm F_{\text{BB},m}$, the objective is upper-bounded by
\vspace{-0.2cm}
\begin{equation}
\begin{aligned}
    \sum_{m=1}^{M}\|\bm F_{m}-\bm F_{\text{RF}}\bm F_{\text{BB},m}\|_{\text{F}}^{2}&=\!\sum_{m=1}^{M}\|(\bm F_{m}\bm F_{\text{BB},m}^{-1}-\bm F_{\text{RF}})\bm F_{\text{BB},m}\|_{\text{F}}^{2} \\
    &\leqslant \! \sum_{m=1}^{M} \|\bm F_{m}\bm F_{\text{BB},m}^{-1}-\bm F_{\text{RF}}\|_{\text{F}}^{2}\|\bm F_{\text{BB},m}\|_{\text{F}}^{2},
\end{aligned}
\end{equation}
according to Cauchy-Schwarz inequality. Then $\bm F_{\text{RF}}$ and $\bm F_{\text{BB},m}$ are decoupled. The design problem in terms of $\bm F_{\text{RF}}$ can be transformed into 
\vspace{-0.35cm}
\begin{align}
    \underset{\bm F_{\text{RF}}}{\text{min}}&~\sum_{m=1}^{M} \|\bm F_{m}\bm F_{\text{BB},m}^{-1}-\bm F_{\text{RF}}\|_{\text{F}}^{2}\|\bm F_{\text{BB},m}\|_{\text{F}}^{2} \label{Problem7}\\
    \text{s.t.} &~ \vert \bm F_{\text{RF}}[i,j]  \vert=1. \tag{\ref{Problem7}{a}} \label{Problem7_a}
\end{align}
\vspace{-0.3cm}
The optimal solution can be readily obtained as
\vspace{-0.0cm}
\begin{equation}
    \bm F_{\text{RF}}=\text{exp}\{ j\angle (\sum_{m=1}^{M}\|\bm F_{\text{BB},m}\|_{\text{F}}^{2}\bm F_{m}\bm F_{\text{BB},m}^{-1} ) \}.
    \label{equ:update_analog}
    \vspace{-0.2cm}
\end{equation}

\subsubsection{Optimizing $\bm F_{\text{BB},m}$}
With $\bm F_{\text{RF}}$ fixed, the digital-part optimization problem equals to
\vspace{-0.2cm}
\begin{align}
    \underset{ p_{u,m} }{\text{min}}&~\sum_{m=1}^{M}\sum_{u=1}^{U}\|\bm f_{m,u}-p_{m,u}\bm f_{\text{RF},u}\|_{2}^{2} \label{Problem8}\\
    \text{s.t.} &~ \sum_{m=1}^{M}\sum_{u=1}^{U}p_{m,u}^{2}=\frac{1}{N_{\text{t}}}, \tag{\ref{Problem8}{a}} \label{Problem8_a}
\end{align}
where $\bm f_{\text{RF},u}=\bm F_{\text{RF}}[:,u]$.
By briefly dropping the power constraint, $p_{m,u}$ has the close-form solution through least square (LS) method as
\vspace{-0.2cm}
\begin{equation}
    \hat{p}_{m,u}=\bm f_{\text{RF},u}^{\dagger}\bm f_{m,u},
    \vspace{-0.2cm}
\end{equation}
where $\bm f_{\text{RF},u}^{\dagger}=\frac{1}{N_{\text{t}}}\bm f_{\text{RF},u}^{\text{H}}$ is the pseudo inverse of $\bm f_{\text{RF},u}$.
Taking into account power budget, $\hat{p}_{m,u}$ can be further normalized as
\vspace{-0.2cm}
\begin{equation}
    p_{m,u}=\frac{1}{\sqrt{\sum_{m=1}^{M}\sum_{u=1}^{U} \hat{p}_{m,u}^{2}} } \hat{p}_{m,u}.
    \label{equ:update_digital}
    \vspace{-0.2cm}
\end{equation}

\subsubsection{Iteration}

The update of $\bm F_{\text{RF}}$ and $\bm F_{\text{BB},m}$ is performed alternatively until the maximal number of iterations $N_{\text{iter}}$ is reached. 
Then $\bm F_{\text{BB}}$ and $\bm F_{\text{RF}}$ are obtained at each subframe, for time-averaged utility maximization.
The pseudo code of the proposed optimization algorithm is illustrated in Algorithm~\ref{Alg_1}.

\begin{algorithm}[h]
  \caption{Optimization-based Wideband ISAC Precoding}
  \textbf{Input}:$\overline{\bm H}$, $\theta_{\text{s}}$, $P_{\text{t}}$, $\sigma_{\text{c}}$, $\sigma_{\text{s}}$;\\
  \textbf{Output}:$\bm F_{\text{RF}}$ and $\bm F_{\text{BB},m}$; \\
  \textbf{Steps}:
  \begin{algorithmic}[1]
  \State Compute the optimal solution $\{\bm W_{m,u}^{*}\}$ by solving (\ref{Problem6}) with convex optimization solvers;
  \State Compute $\{\bm f_{m,u}\}$ to construct $\{\bm F_{m}^{*}\}$;
  \State Initialize $\bm F_{\text{BB},m}=\bm I_{\text{U}}$;
  \For{$n=1$ to $N_{\text{iter}}$}
  \State Update $\bm F_{\text{RF}}$ as (\ref{equ:update_analog});
  \State Update $\bm F_{\text{BB},m}$ as (\ref{equ:update_digital});
  \EndFor
  \end{algorithmic}
  \label{Alg_1}
\end{algorithm}
\vspace{-0.3cm}

\subsection{Remark on optimization-based schemes}
The proposed optimization-based algorithm can serve as a candidate solution for the wideband ISAC precoding. However, it is far away from optimal in practical double dynamics, due to the following reasons:
\begin{itemize}
    \item \textbf{Huge complexity}: The iteration in the solution introduces significant computational complexity, scaling up to $\mathcal{O}(M\!(N_{\text{t}}^{3.5}\!\log(1/\epsilon)\!\!+\!\!2N_{\text{iter}}U^{2}N_{\text{t}}))$ given a solution accuracy $\epsilon$ and $N_{\text{iter}}$ alternating iterations. This results in an unacceptable delay, especially with large antenna arrays in studied scenarios.
    \item \textbf{Heavy reliance on perfect prior}: The doubly-dynamic scenarios impose significant challenges on instantaneous channel estimation and targets' position acquisition,
    rendering the optimization-based method, which heavily rely on perfect state information, no longer effective.
    \item \textbf{Mediocre performance}: 
    The obtained solutions may incur potentially significant suboptimality for two main reasons. Firstly, the relaxation of the rank-1 constraint can lead to being stuck in local optimality when optimizing $\bm F_{m}$. Secondly, the decomposition of $\bm F_{m}$ introduces additional loss in overall performance.
\end{itemize}

Therefore, it is crucial to solve this sequential optimization problem with low complexity and reduced reliance on perfect instantaneous state information.

\begin{figure*}[t]
  \vspace{-0.6cm}
  \centering
  \includegraphics[width=0.63\linewidth]{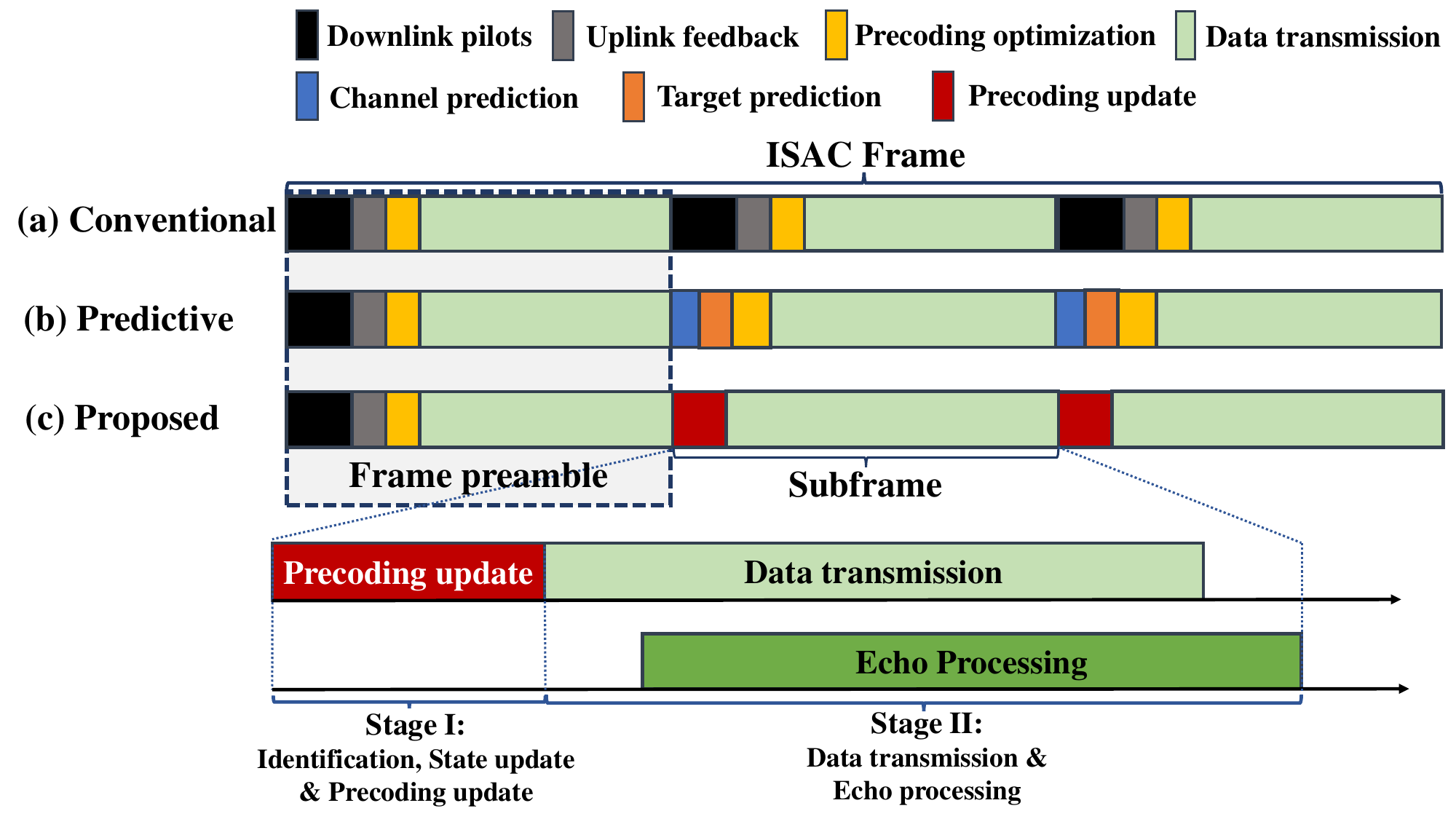}
  \vspace{-0.1cm}
  \captionsetup{justification=centering}
  \captionsetup{font={small}}
  \caption{The frame structure of conventional, predictive, and proposed one for ISAC.}
  \vspace{-0.4cm}
  \label{fig:frame_structure}
\end{figure*}

\section{Frame structure for ISAC processing in double dynamics}

In this section, we introduce a dedicated frame structure to facilitate ISAC operations in double dynamics.
Instead of relying solely on instantaneous environmental prior information, the dual-functional waveform possesses inherent sensing capabilities. 
Taking inspiration from this, the update of hybrid precoding is performed in advance based on historical observations from BS itself. The ISAC frame structure is illustrated in Fig.~\ref{fig:frame_structure}, where each frame begins with a frame preamble, followed by $T$ subframes containing $L$ OFDM symbols.

In the frame preamble, initial CSI can be obtained using existing channel estimation techniques. Simultaneously, an approximate prior on the target's angle can be estimated through omni-directional angle scanning. Utilizing this initial state, the optimization-based hybrid precoding procedure outlined in Algorithm 1 is implemented in the frame preamble for data transmission. In the subsequent subframes, multi-user communication occurs alongside mono-static target sensing conducted by the BS. To elaborate further, each subframe consists of two stages: Precoding update and data transmission.

\subsubsection{Stage \RNum{1}: \!Precoding update}
\!In this stage, historical estimates from the ISAC BS are utilized instead of acquiring real-time environmental state. Using the proactively-obtained estimates, the BS updates the observed state, enabling the hybrid precoders to be updated according to the current strategy.

\subsubsection{Stage \RNum{2}: Data transmission}   
With $\bm F_{\text{RF}}$ and $\bm F_{\text{BB}}$ updated at the beginning of each subframe, the data signal is precoded and then transmitted to both users and the target. The echo is processed at BS for target estimation via existing algorithms. We assume that the transmitted signal does not interfere with the reception antennas, as this issue can be mitigated through various full-duplex radio techniques \cite{full_duplex_ISAC}. These target estimates are then utilized for state update in the following subframes, creating a closed loop of sensing information.


Various frame structures that can be utilized in doubly-dynamic ISAC scenarios are compared, as depicted in Fig.~\ref{fig:frame_structure}.
The operation in \cite{protocol_a} conducts the precoding design based on the real-time channel estimation and target detection through downlink pilots and uplink feedback in each subframe. This leads to a decrease in SE due to high time overhead and computational burden, making it unsuitable for double dynamics.
In contrast, the frame structure in \cite{protocol_b, dynamic_MP} proactively predicts CSI and the target's position to minimize the time overhead associated with prior acquisition. However, real-time optimization of precoding remains crucial, and the computational complexity remains essential. The proposed ISAC frame structure alleviates the dependence on real-time prior acquisition, significantly reducing time overhead. Moreover, precoding updates are performed without the need for complete optimization, thus lowering computational complexity.

\section{DRL-aided ISAC precoding in double dynamics}

Based on the ISAC frame structure for doubly-dynamic scenarios in Section IV, we propose a DRL-aided approach for ISAC precoding with hybrid precoding structure in this section. We first formulate the problem as a POMDP, where the ISAC BS acts as an agent and explores the environment to gain experiences for refining its precoding strategies.

When the environment state is partially observed at subframe $t$, BS takes an action according to its strategy. Subsequently, the agent receives a reward and the environment evolves to the next state. This process continues, with the new state being observed at each step, and new actions are taken accordingly. The key components are detailed in this section.

\vspace{-0.3cm}
\subsection{Environmental State Space}

The environmental state is expressed as
\vspace{-0.2cm}
\begin{equation}
    \bm S^{(n)}\triangleq\left\{\mathcal{H}^{(n)}, \mathcal{U}^{(n)}, \mathcal{T}^{(n)} \right\} \in \mathcal{S},
\vspace{-0.2cm}
\end{equation}
where $\mathcal{H}^{(n)}$, $\mathcal{U}^{(n)}$ and $\mathcal{T}^{(n)}$ denote the instantaneous state of channel information, users' positions and the target's position, respectively.
$\mathcal{H}^{(n)}=[ \bm H_{0}^{(n)}, \cdots, \bm H_{N_{d}-1}^{(n)} ]$ contains the time-domain channel in $N_{\text{d}}$ delay taps.
$\mathcal{U}^{(n)}=\{ (\theta_{u}^{(n)}, d_{u}^{(n)})\vert_{u=1}^{U} \}$ and $\mathcal{T}^{(n)}=\{(\theta_{\text{s}}^{(n)}, d_{\text{s}}^{(n)})\}$ contain the polar coordinates of the users and of the target in the $n$-th subframe, respectively.

\setlength{\parindent}{0pt}
\textbf{BS-End Observation:}
\setlength{\parindent}{10pt}

In doubly-dynamic scenarios, BS obtains part of the environmental state, which can be expressed as
\vspace{-0.2cm}
\begin{equation}
    \mathcal{O}(\bm S^{(n)})=\left\{ \bm S_{\text{H}}^{(n)}, \bm S_{\text{P}}^{(n)}  \right\},
    \vspace{-0.2cm}
\end{equation}
where $\bm S_{\text{H}}^{(n)}$ and $\bm S_{\text{P}}^{(n)}$ denote the channel-related and the position-related observations.

\subsubsection{Channel-related Observation}
Considering that the instantaneous CSI at each subframe is prohibitive in fast-fading channels, only the initial CSI is estimated at each frame preamble. The initial CSI is pre-processed as 
\vspace{-0.1cm}
\begin{equation}
    \bm S_{\text{H}}^{(n)}\!=\![\vert \text{vec}(\widehat{\bm H}_{0}^{(0)} \bm D_{\text{t}})\vert,\! \vert\text{vec}(\widehat{\bm H}_{1}^{(0)} \bm D_{\text{t}})\vert, \!\cdots,\!\vert\text{vec}(\widehat{\bm H}_{N_{d}\!-\!1}^{(0)} \bm D_{\text{t}})\vert],
\end{equation}
where $\widehat{\bm H}_{d}^{(0)}$ is the estimate of the $d$-th tap CSI $\widetilde{\bm H}_{d}^{(0)}(0)$ at the beginning of each frame ($n=0$), $\vert(\cdot)\vert$ denotes the modulo operation, and $\bm D_{\text{t}}\in \mathbb{C}^{N_{\text{t}}\times G_{\text{t}}}$ is the angular dictionary defined in \cite{DSDS_TWC} with size $G_{t}$, to transform initial CSI into a sparser representation in beamspace and delay domain.

\begin{table}[t]
    \centering
    \caption{Notations related to the environmental observations.}
    \begin{tabular}{|p{2.4cm}||p{5.6cm}|}
    \hline
       Notation  &  Definition \\
       \hline
       \hline
       $\bm S_{\text{H}}^{(n)}$ & Channel-related observations as network input \\
       \hline
       $\bm S_{\text{P}}^{(n)}$ & Position-related observations as network input \\
       \hline
       $\widehat{\bm H}_{d}^{(0)}$  &  Estimate of the $d$-th tap of initial CSI\\
       \hline
       $\bm D_{\text{t}}$  &  Beamspace dictionary \\
        \hline
       $\mathcal{P}_{\text{ob}}^{(n)}$  &  The set of coordinates of user-target estimates  \\
       \hline
       $\mathcal{U}_{\text{ob}}^{(n)}$  &  The BS-end estimates of the users' positions  \\
       \hline
       $\mathcal{T}_{\text{ob}}^{(n)}$  &  The BS-end estimates of the target's position \\
       \hline
       $ ({\theta_{i}^{\text{ob}}}^{(n)}, {d_{i}^{\text{ob}}}^{(n)})$ &  BS-end estimation of the $i$-th object's coordinates \\
       \hline
       $ ({\theta_{u}^{\text{fb}}}^{(n)}, {d_{u}^{\text{fb}}}^{(n)})$  &  Periodic GPS feedback from the $u$-th user \\
       \hline
    \end{tabular}
    \label{tab:notation}
    \vspace{-0.4cm}
\end{table}

\subsubsection{Position-related Observation}
The position-related observation includes the BS-end position estimates of both users and the target, denoted by
\begin{equation}
\begin{aligned}
    \mathcal{U}_{\text{ob}}^{(n)}&=\{ {(\hat{\theta}_{u}^{(n)}, \hat{d}_{u}^{(n)})\vert _{u=1}^{U} }\}, \\
    \mathcal{T}_{\text{ob}}^{(n)}&=\{ (\hat{\theta}_{s}^{(n)}, \hat{d}_{s}^{(n)}) \},
\end{aligned}
\end{equation}
where $(\hat{\theta}_{u}^{(n)}, \hat{d}_{u}^{(n)})$ and $(\hat{\theta}_{s}^{(n)}, \hat{d}_{s}^{(n)})$ are the estimated coordinates of the users and the target respectively.
Denote $\mathcal{P}_{\text{ob}}^{(n)}=\mathcal{U}_{\text{ob}}^{(n)} \bigcup \mathcal{T}_{\text{ob}}^{(n)}=  \{ ({\theta_{i}^{\text{ob}}}^{(n)}, {d_{i}^{\text{ob}}}^{(n)}) \vert_{i=1}^{U+1} \}$ as the set of the estimated coordinates of the objects through echo processing in the $n$-th subframe.
Note that BS itself lacks the ability to differentiate between $\mathcal{U}_{\text{ob}}^{(n)}$ and $\mathcal{T}_{\text{ob}}^{(n)}$ from $\mathcal{P}_{\text{ob}}^{(n)}$. 
In fact, users can periodically upload their locations from GPS to the BS every $T_{\text{fb}}$ subframes, as operated in \cite{location_aware_overview}. These feedbacks are denoted by $\{ ({\theta_{u}^{\text{fb}}}^{(n)}, {d_{u}^{\text{fb}}}^{(n)})_{u=1}^{U} \}$. In subframes with the real-time uploaded feedback, the BS differentiates $U$ users from $\mathcal{P}_{\text{ob}}^{(n)}$ in sequence, based on the nearest neighbor criterion:
\vspace{-0.2cm}
\begin{equation}
    \{{\hat{\theta}_{u}^{(n)}, \hat{d}_{u}^{(n)} }\}\!=\!\! \underset{  {\theta_{i}^{\text{ob}}}^{(n)}, {d_{i}^{\text{ob}}}^{(n)}}{\text{argmin}} \left [\frac{{\theta_{i}^{\text{ob}}}^{(n)}\!\!-\!{\theta_{u}^{\text{fb}}}^{(n)}}{\pi}\right ]^{2}\!\!+\!\left [ \frac{{d_{i}^{\text{ob}}}^{(n)}\!\!-\!{d_{u}^{\text{fb}}}^{(n)}}{d_{\text{max}}}\right ]^{2}.
\end{equation}
Then the remaining element in $\mathcal{P}_{\text{ob}}^{(n)}$ is the target's position estimate, composing $\mathcal{T}_{\text{ob}}^{(n)}$. In subframes without feedbacks, BS identifies the target and users based on their positions in the last subframe, since they exhibit slight movement across subframes. 
With the target and users identified in $\mathcal{T}_{\text{ob}}$ and $\mathcal{U}_{\text{ob}}$ per subframe, we represent the observation using a three-dimensional time-angle-range spectrum, $\bm S_{\text{P}}^{(n)}$, i.e., 
\vspace{-0.2cm}
\begin{equation}
    \bm S_{\text{P}}^{(n)}=\left\{\bm P^{(n-3)}, \bm P^{(n-2)}, \bm P^{(n-1)} \right\},
    \vspace{-0.2cm}
\end{equation}
where $\bm P^{(n)}\in \mathbb{R}^{N_{\text{x}}\times N_{\text{y}}}$ denotes the estimated angle-range spectrum at $n$ divided into $N_{\text{x}}\times N_{\text{y}}$ grids, with a sensing range $[\theta_{\text{min}}, \theta_{\text{max}}) \times [d_{\text{min}}, d_{\text{max}})$. Specifically,
\vspace{-0.2cm}
\begin{equation}
  \bm P^{(n)}(n_{\text{x}},\! n_{\text{y}})\! = \!\!
  \begin{cases}
    1, \!\!&\text{if $\{[\theta_{n_{\text{x}}},\!\theta_{n_{\text{x}}\!+\!1})\!\times \! [d_{n_{\text{y}}},\! d_{n_{\text{y}}\!+\!1})\}\! \bigcap \mathcal{U}_{\text{ob}}^{(n)} \!\!\neq \!\emptyset$},\\
	2,\!\! &\text{if $\{[\theta_{n_{\text{x}}},\!\theta_{n_{\text{x}}\!+\!1})\!\times \! [d_{n_{\text{y}}},\! d_{n_{\text{y}}\!+\!1})\}\! \bigcap \mathcal{T}_{\text{ob}}^{(n)} \!\!\neq \!\emptyset$},\\
    0,\!\! &\text{Otherwise},
  \end{cases}
  \vspace{-0.2cm}
\end{equation}
where $\theta_{n_{\text{x}}}\!=\!\theta_{\text{min}}\!+\!\frac{\theta_{\text{max}}\!-\!\theta_{\text{min}}}{N_{\text{x}}}(n_{\text{x}}-1)$ and $d_{n_{\text{y}}}\!=\!d_{\text{min}}\!+\frac{d_{\text{max}}-d_{\text{min}}}{N_{\text{y}}}(n_{\text{y}}-1)$.  It is worth noting that this representation can adapt to varying numbers of users and targets.

\vspace{-0.3cm}
\subsection{Hybrid Action Space}
\vspace{-0.1cm}

\begin{figure}[t]
  \vspace{-0.3cm}
  \centering
 \includegraphics[width=0.9\linewidth]{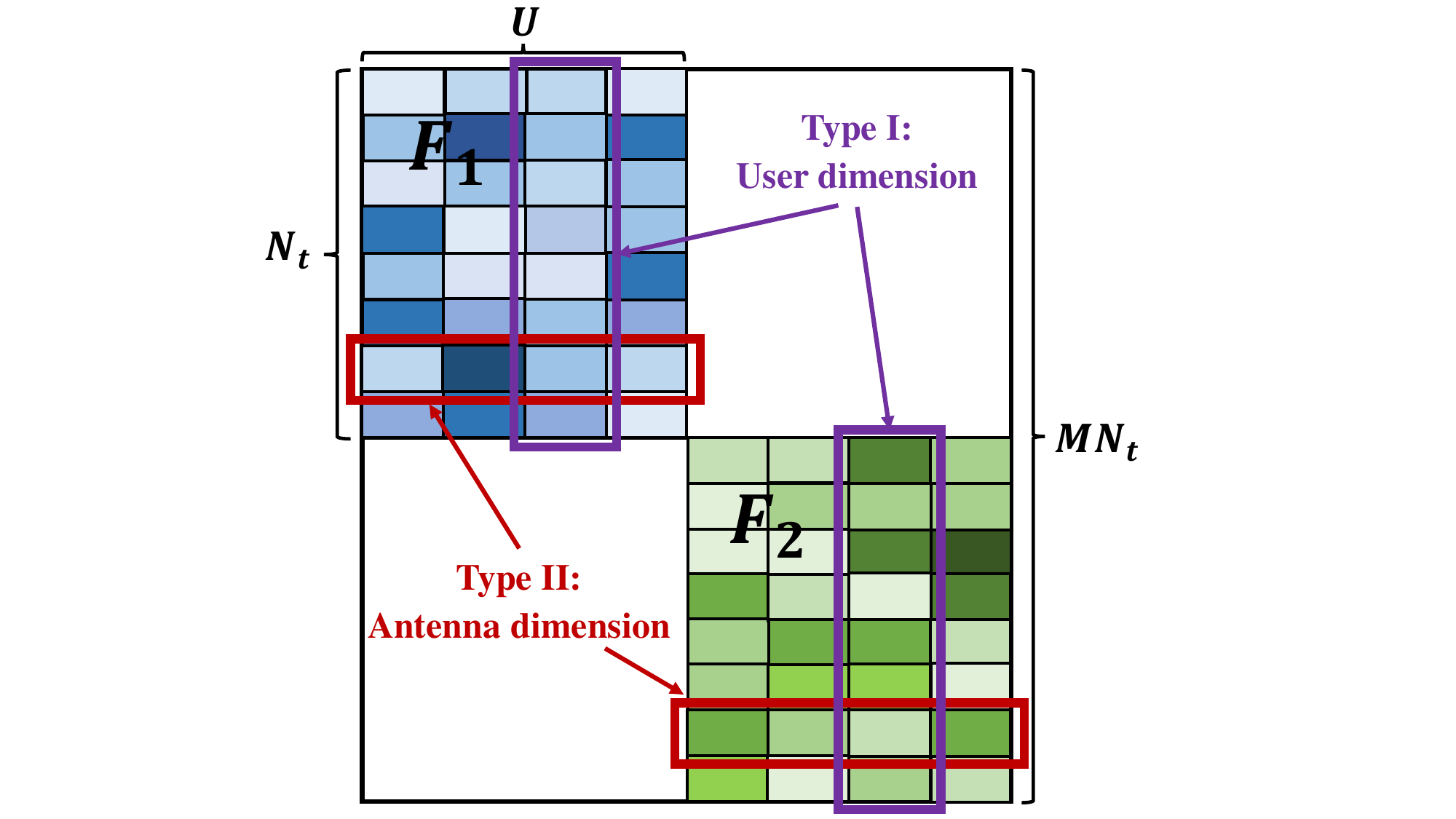}
  \vspace{-0.2cm}
  \captionsetup{font={small}}
  \caption{An illustration of two types of precoding update schemes ($N_{\text{t}}=8, U=4, M=2$).}
  \vspace{-0.6cm}
  \label{fig:precoding_update}
\end{figure}

Given that the environment typically does not undergo drastic changes between adjacent subframes, we suggest updating $\bm F_{\text{RF}}$ and $\bm F_{\text{BB}}$ progressively, instead of optimizing them from scratch in each subframe.
To effectively update discrete phase shifters and continuous digital precoding simultaneously, we propose to update $\bm F_{\text{RF}}$ and $\bm F_{\text{BB}}$ in hybrid action spaces, including two operation types, i.e., update across user dimension and across antenna dimension respectively, as illustrated in Fig.~\ref{fig:precoding_update}.

\setlength{\parindent}{0pt}
\textbf{Type \RNum{1}: Update Across User Dimension}

\setlength{\parindent}{10pt}

Inspired from the progressive update in \cite{GC}, we propose an incremental updating scheme across the user dimension. In each subframe, one of the columns in $\bm F_{\text{RF}}$ is selected to be updated, followed by updating the corresponding parameters in $\bm F_{\text{BB}}$.
The element-wise updating is operated in each column independently, corresponding to each user.

\subsubsection{Action Space Design}
The hybrid action combination can be expressed as $\bm A^{(n)}=\{ \bm a^{(n)}_{\text{RF},1}, \bm a^{(n)}_{\text{RF},2}, \bm a^{(n)}_{\text{BB}} \}$.
$\bm a_{\text{RF},1}^{(n)}\in [0,1]^{\text{U}^{\text{max}}}$ represents for the probability distribution of the $U$ columns in $\bm F_{\text{RF}}^{(n-1)}$ to be selected, and $\bm a_{\text{RF},2}^{(n)}\in [0,1]^{N_{\text{t}
}}$ represents for the incremental update of the $N_{\text{t}}$ elements in the selected column. $\bm a^{(n)}_{\text{BB}}\in  [-1,1]^{M}$ denotes the increment of $p_{m,u}$ in $\bm F_{\text{BB},m}$ ($m \in [1,M]$).

\subsubsection{Precoding Update}
BS First chooses one column in $\bm F_{\text{RF}}^{(n-1)}$ according to the probability distribution represented by $\bm a_{\text{RF},1}^{(n)}$. Denote $u^{*}$ as the index of the selected column in $\bm F_{\text{RF}}$, i.e.,
\vspace{-0.3cm}
\begin{equation}
    \mathbb{P}(u^{*}=u)=\frac{\bm a_{\text{RF},1}^{(n)}[u]}{\|\bm a_{\text{RF},1}^{(n)}\|_{1}}.
\vspace{-0.2cm}
\end{equation}
Then element-wise incremental update is performed on the $u^{*}$-th column as
\vspace{-0.25cm}
\begin{equation}
    \angle \bm F^{(n)}_{\text{RF}}[i,u^{*}]\!=\!
    \begin{cases}
        \angle \bm F^{(n-1)}_{\text{RF}}[i,u^{*}] \!+\! \Delta \theta, &\text{if $\bm a_{\text{RF},2}^{(n)}[i]\!\geqslant \!0 $},\\
        \angle \bm F^{(n-1)}_{\text{RF}}[i,u^{*}] \!-\! \Delta \theta, &\text{if $ \bm a_{\text{RF},2}^{(n)}[i] \!< \!0 $},
    \end{cases}
    \vspace{-0.2cm}
\end{equation}
for $\forall i\in [1, N_{\text{t}}]$,
and $\Delta \theta=\frac{2\pi}{2^{B}}$.
The digital precoder corresponding to the $u^{*}$-th user is updated as
\begin{equation}
\vspace{-0.2cm}
    p^{(n)}_{m,u^{*}}\!=\!p^{(n-1)}_{m,u^{*}}\!+\!\bm a^{(n)}_{\text{BB}}[m],~~ m \!\in \![1,M].
\end{equation}
Then $\bm F^{(n)}_{\text{BB},m}$ is normalized to satisfy the power constraint.
\\

\setlength{\parindent}{0pt}
\textbf{Type \RNum{2}: Update Across Antenna Dimension}
\setlength{\parindent}{10pt}

Taking a step further, we introduce an additional updating strategy involving the antenna dimension. In each subframe, a single antenna is chosen, and all precoding parameters associated with that antenna are updated. This approach ensures that the size of the action space remains constant regardless of the number of users.

\setcounter{subsubsection}{0}
\subsubsection{Action Space Design}
In this updating scheme, the action is defined as $\bm A^{(n)}=[\bm a_{\text{RF}}^{(n)}, \bm a_{\text{BB}}^{(n)} ]$. $\bm a_{\text{RF}}^{(n)} \in [0,1]^{N_{\text{t}}}$ denotes the probability distribution of the $N_{\text{t}}$ antennas being selected to be updated in the $n$-th subframe. $\bm a_{\text{BB}}^{(n)}\in [-1,1]^{M}$ denotes the phases of the differentials at $M$ subcarriers for digital update.

\subsubsection{Precoding Update}

Firstly, one of the $N_{\text{t}}$ antennas is selected based on the probability given by $\bm a_{\text{RF}}^{(n)}$. Denote $i^{*}$ as the index of the selected antenna, i.e., 
\vspace{-0.2cm}
\begin{equation}
    \mathbb{P}(i^{*}=i)=\frac{\bm a_{\text{RF}}^{(n)}[i]}{\|\bm a_{\text{RF}}^{(n)}\|_{1}}.
    \vspace{-0.2cm}
\end{equation}
Subsequently the hybrid precoding matrices corresponding to the $i^{*}$-th antenna, i.e., $\bm F_{m}[i^{*},:]$ ($m\in [1,M]$) are updated simultaneously based on $\bm a_{\text{BB}}^{(n)}$.
Specifically, the differential for $\bm F_{m}[i^{*}, u]$ is
\vspace{-0.3cm}
\begin{equation}
    g^{(n)}_{m,u}=\vert \bm F_{m}^{(n-1)}[i^{*}, u] \vert e^{j\pi\angle(\bm a^{(n)}_{\text{BB}}[m])}.
    \vspace{-0.2cm}
\end{equation}
Then the element-wise update for $\bm F_{m}[i^{*},:]$ ($m\in [1,M]$) is performed as
\vspace{-0.25cm}
\begin{equation}
    \bm F_{m}^{(n)}[i^{*}, u]=\bm F_{m}^{(n-1)}[i^{*}, u]+\eta_{\text{r}} g_{u,m}^{(n)}, \forall u\in [1,U],
    \vspace{-0.2cm}
\end{equation}
where $\eta_{\text{r}}$ represents the updating rate for the differentials. Both the amplitudes and the phases corresponding to the $i^{*}$-th antenna in the hybrid precoding are updated as
\vspace{-0.25cm}
\begin{equation}
\begin{aligned}
    &\angle \left \{\bm F_{\text{RF}}^{(n)}[i^{*},u]\right \}=\angle \left \{\bm F_{m}^{(n)}[i^{*}, u]\right \},  \\
    &\vert \bm F_{\text{BB},m}^{(n)}[u,u] \vert = \vert \bm F_{m}^{(n)}[i^{*}, u] \vert.
\end{aligned}
\vspace{-0.3cm}
\end{equation}
Subsequently the power normalization of $\bm F_{\text{BB},m}^{(n)}$ is performed,
and the analog phases in $\bm F_{\text{RF}}^{(n)}$ is quantized to the finite-resolution set according to the nearest neighbor criterion, i.e.,
\vspace{-0.3cm}
\begin{equation}
    \mathcal{Q}(\angle \bm F^{(n)}_{\text{RF}}[i^{*},u]) 
    =\mathop{\arg\min}\limits_{\theta \in \mathcal{B}} \vert e^{j\theta}-e^{j\angle \bm F^{(n)}_{\text{RF}}[i^{*},u]} \vert^{2}.
\vspace{-0.2cm}
\end{equation}
\vspace{-0.2cm}

\begin{remark}
\vspace{-0.2cm}
The updates across the user dimension are compatible with the multi-beam structure (MBS) using lens arrays, while updates across the antenna dimension are suitable for multi-aperture structures (MAS) with PS networks. Therefore, the proposed updating schemes can adapt to existing hybrid precoding hardware architectures.
\end{remark}

\vspace{-0.4cm}
\subsection{Reward Function}

To guide the BS to maximize the long-term ISAC utility in double-dynamic scenarios, we design the immediate reward function as
\vspace{-0.40cm}
\begin{equation}
    r^{(n)}=\psi \frac{\mathcal{R}^{(n)}}{{\mathcal{R}^{*}}^{(n)}} + (1-\psi) \frac{\mathcal{J}(\theta_{\text{s}}^{(n)}) } { \mathcal{J}^{*}(\theta_{\text{s}}^{(n)})} .
\vspace{-0.2cm}
\end{equation}
Following the strategy parameterized by $\mu$, the long-term discounted reward starting from $n_{0}$ is
\vspace{-0.25cm}
\begin{equation}
    R(\mu)=\mathbb{E}_{\mu}[\sum_{n=n_{0}}^{T}\gamma^{n-n_{0}}r^{(n)}],
\vspace{-0.2cm}
\end{equation}
where $\gamma$ denotes the discounting factor. The objective of BS in this POMDP is to find
\vspace{-0.3cm}
\begin{equation}
     \mu^{*}=\underset{ \mu }{\text{argmax}} ~R(\mu). \label{Problem9} 
     \vspace{-0.2cm}
\end{equation}

\vspace{-0.3cm}
\subsection{State transition}
\begin{figure}[b]
  \vspace{-0.6cm}
  \centering
  \includegraphics[width=0.90\linewidth]{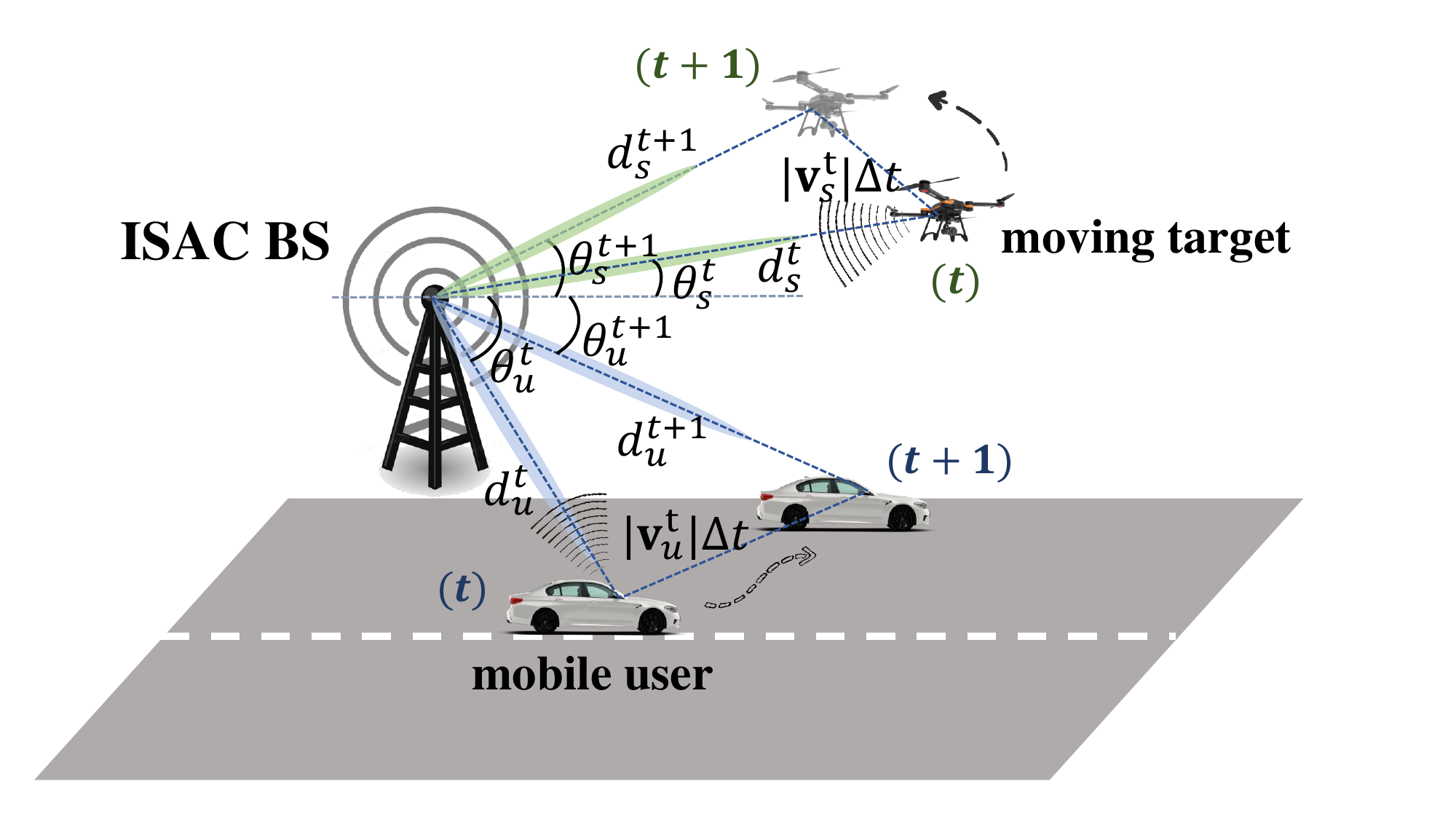}
  \vspace{-0.2cm}
  \captionsetup{font={small}}
  \caption{An illustration of the environmental state transition.}
  \vspace{-0.6cm}
  \label{fig:state transition}
\end{figure}

In the $(n+1)$-th subframe, the environmental state is evolved into $\bm S^{(n+1)}=\{\mathcal{H}^{(n+1)}, \mathcal{U}^{(n+1)}, \mathcal{T}^{(n+1)} \}$. Denote $\{\bm v^{(n)}_{u}\}_{u=1}^{U}$ and $\bm v^{(n)}_{s}$ as the velocity of $U$ users and of the target in the $n$-th subframe respectively. The environmental state varies according to these current velocities, as demonstrated in Fig.~\ref{fig:state transition}
\subsubsection{Transition of $\mathcal{U}^{(n)}$ and $\mathcal{T}^{(n)}$}

The positions of users $\mathcal{U}^{(n+1)}=\{ (\theta_{u}^{(n+1)}, d_{u}^{(n+1)})\vert_{u=1}^{U} \}$ varies according to $\bm v_{u}$, and the specific evolution process is characterized by kinematic equations as
\vspace{-0.2cm}
\begin{align}
    &\sin{(\theta_{u}^{(n+1)}\!-\!\theta_{u}^{(n)})}=\vert \bm v_{u}^{(n)}\vert \Delta T \sin{(\theta_{u}^{(n)}\!-\!\angle(\bm v_{u}^{(n)})\!+\!\pi)}/d_{u}^{(n+1)}, \nonumber \\
    &(d_{u}^{(n+1)})^{2}\!-\!(d_{u}^{(n)})^{2}=(\vert \bm v_{u}^{(n)}\vert \Delta T)^{2} \nonumber \\
    &~~~~~~~~~~~-2d_{u}^{(n)}\vert \bm v_{u}^{(n)}\vert \Delta T \cos{(\theta_{u}^{(n)}-\angle(\bm v_{u}^{(n)})\!+\!\pi)}. 
    \vspace{-0.2cm}
     \label{equ:transition}
\end{align}
The target position is $\mathcal{T}^{(n+1)}=\{(\theta_{\text{s}}^{(n+1)}, d_{\text{s}}^{(n+1)})\}$, evolving similarly  according to $\bm v_{\text{s}}^{(n)}$.
It can be observed that the positions at $(t+1)$ is determined by the positions and the velocities at $t$. Thus the evolutions of $\mathcal{U}^{(n)}$ and $\mathcal{T}^{(n)}$ are Markov chains.

\subsubsection{Transition of $\mathcal{H}^{(n)}$}

The channel state in the $(n+1)$-th subframe is $\mathcal{H}^{(n+1)}=[ \bm H_{0}^{(n+1)}, \cdots, \bm H_{N_{d}-1}^{(n+1)} ]$. The parameters in $\mathcal{H}$ evolves as
\begin{equation}
    \begin{aligned}
        (\theta_{u}^{(n+1)}, d_{u}^{(n+1)})&\in \mathcal{U}^{(n+1)},\\
        \tau_{u}^{(n+1)}&=2 d_{u}^{(n+1)}/c_v, \\
        \omega_{u}^{(n+1)}&=2\pi f_{c}\vert \bm v_{u}\vert T_{s}\sin{\theta_{u}^{(n+1)}}/c_v,\\
        \beta_{u}^{(n+1)}&=\beta_{u}^{(n)}+ \Delta \beta,
    \end{aligned}
    \vspace{-0.2cm}
\end{equation}
where $\Delta \beta\sim \mathcal{CN}(0, \sigma_{\Delta \beta})^{2}$ denote the variation of the complex gain of the paths. The parameters of channels satisfy the Markov property, meaning that $\mathcal{H}^{(n)}$ also forms a Markov chain.

At BS, the channel-related observation $\bm S_{\text{H}}^{(n+1)}=\bm S_{\text{H}}^{(n+1)}$ remains the initial estimate, while the position-related observation $\bm S_{\text{P}}^{(n+1)}=\left\{ \bm P^{(n-2)}, \bm P^{(n-1)}, \bm P^{(n)}\right\}$ is updated with the latest estimation, $\bm P^{(n)}$, through echo processing in the $n$-th subframe.
It can be verified that the observations also exhibit Markov properties.

\begin{figure*}[t]
  \vspace{-0.0cm}
  \centering
  \includegraphics[width=0.76 \linewidth]{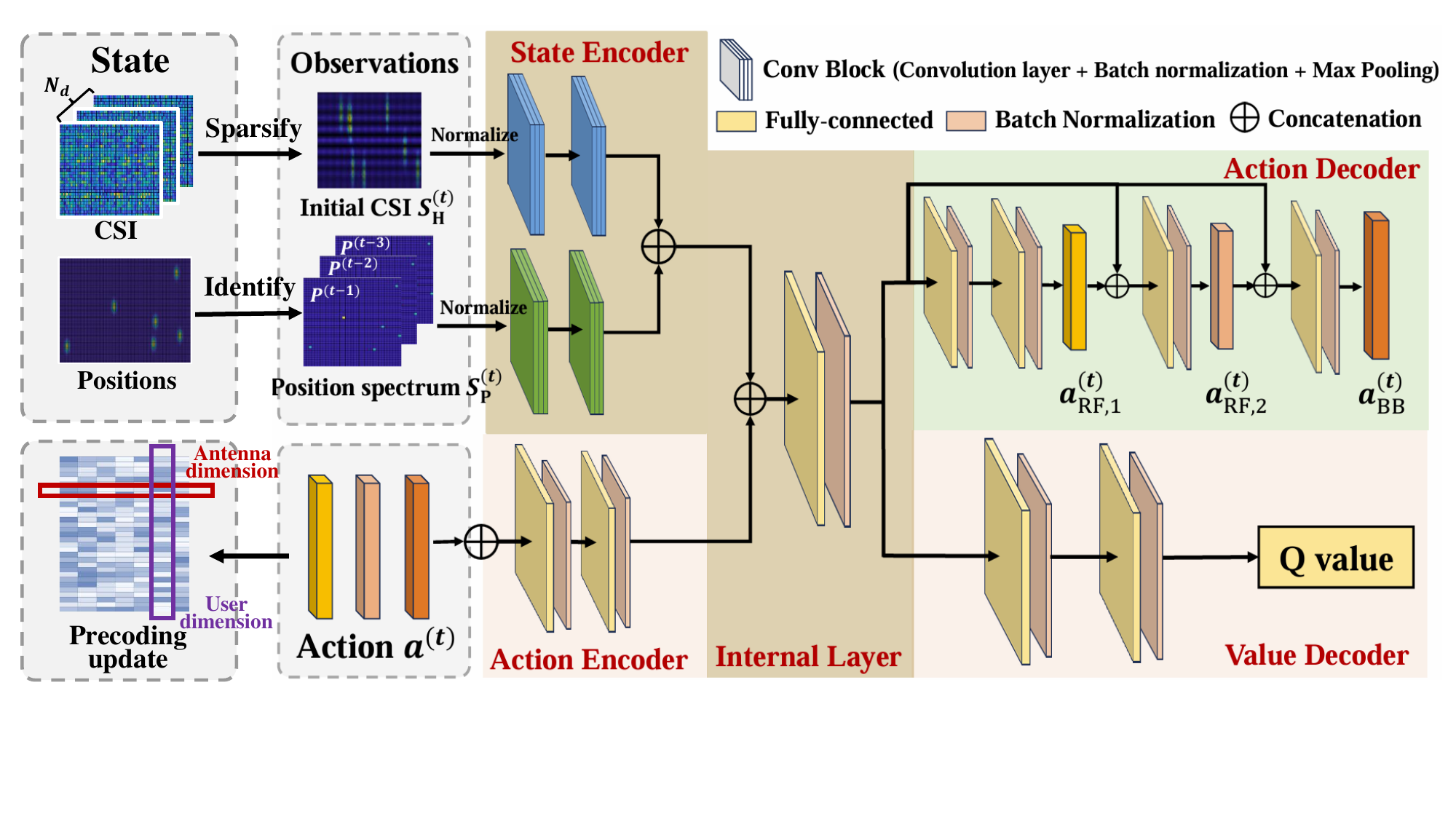}
  \vspace{-1.2cm}
  \captionsetup{font={small}}
  \caption{The PSAC network architecture.}
  \label{fig:network architecture}
  \vspace{-0.4cm}
\end{figure*}

\vspace{-0.3cm}
\subsection{Learning Architecture}

In each subframe, updating both $\bm F_{\text{RF}}$ and $\bm F_{\text{BB},m}$ results in a high-dimensional hybrid action space. Traditional algorithms like Deep Q-Network (DQN) may experience a decline in performance when faced with this challenge \cite{DDPG}. Additionally, the observations display intricate structures, making it difficult to extract features for decision-making. To address these challenges, we utilize an actor-critic framework with a tailored network architecture and specialized training procedures.

Due to the different sizes of $\bm S_{\text{H}}^{(n)}$ and $\bm S_{\text{P}}^{(n)}$, two input headers are required to process them separately. This will result in an increase in the number of parameters for the actor and critic networks, thereby increasing the difficulty of network training. 
It's worth noting that both the actor and critic networks require feature extraction of the state, so we share some of the parameters between them and adopt a parameters-shared actor-critic (PSAC) architecture. 
The overall network architecture comprises five parts: state encoder, action encoder, internal layer, action decoder, and value decoder. In this setup, the state encoder and internal layer are shared by both the actor and critic components, as illustrated in Fig.~\ref{fig:network architecture}.

\subsubsection{State encoder}
The state encoder, parameterized by $\phi_{\text{SE}}$, pre-processes the observation to extract the features for both the actor and the critic, including two parallel streams of convolutional blocks for processing $\bm S_{\text{H}}$ and $\bm S_{\text{P}}$ respectively. Each convolutional block consists of a convolution layer, a batch normalization (BN) layer and a maximum pooling layer, activated by rectified linear units (ReLU).
The output feature is $\bm \upsilon_{\text{s}}=\phi_{\text{SE}}(\bm S_{\text{H}}, \bm S_{\text{P}})$.

\subsubsection{Action encoder}
The action encoder, parameterized by $\phi_{\text{AE}}$, pre-processes the actions for the critic. It is constituted of fully-connected layers, each of which is followed by a BN layer to stabilize the training. The output of the action encoder is $ \bm \upsilon_{\text{a}}=\phi_{\text{AE}}(\bm a_{\text{RF}}, \bm a_{\text{BB}})$.

\subsubsection{Internal layer}
Parameterized by $\phi_{\text{IN}}$, the internal layer concatenates $\bm \upsilon_{\text{s}}$ and $\bm \upsilon_{\text{a}}$, and then processes them for the decoders through one fully-connected layer and one BN layer. The output is $\bm \upsilon=\phi_{\text{IN}}(\bm \upsilon_{\text{s}}, \bm \upsilon_{\text{a}})$.

\subsubsection{Action decoder}
The action decoder, parameterized by $\phi_{\text{AD}}$, is constituted of two modules to cope with the hybrid action, namely, analog precoding updating network (APU-Net) parameterized by $\phi_{\text{APU}}$ and digital precoding updating network (DPU-Net) parameterized by $\phi_{\text{DPU}}$.
APU-Net processes $\bm \upsilon$ from the internal layer and outputs $\bm a_{\text{RF}}$ for analog precoding update, i.e., $\bm a_{\text{RF}}=\phi_{\text{APU}}(\bm \upsilon)$. Then $\bm \upsilon$ and $\bm a_{\text{RF}}$ are input into DPU-Net for digital precoding update and activated by hyperbolic tangent function, i.e., $\bm a_{\text{BB}}=\phi_{\text{DPU}}(\bm \upsilon, \bm a_{\text{RF},1}, \bm a_{\text{RF},2})$.

\subsubsection{Value decoder}
The value decoder, parameterized by $\phi_{\text{VD}}$, assesses the observation-action pair $(\mathcal{O}^{(n)}, \bm A^{(n)})$ to output the Q-value $Q=\phi_{\text{VD}}(\bm \upsilon)$, through fully-connected layers and BN layers, activated by ReLU functions.

In this PSAC,
the actor network $\mu$, parameterized by $\phi_{\text{A}}=\phi_{\text{SE}}\bigcup \phi_{\text{I}}\bigcup \phi_{\text{AD}}$, maps from the state space to an action $\bm A\in \mathcal{A}$ as $\bm A^{(n)}=\mu(\mathcal{O}^{(n)} \vert \phi_{\text{A}})$.
The critic network, parameterized by $\phi_{\text{C}}=\phi_{\text{SE}}\bigcup \phi_{\text{AE}} \bigcup \phi_{\text{IN}}\bigcup \phi_{\text{VD}}$, maps from the joint observation space and action space into reward, playing a role of action evaluation as $Q(\mathcal{O}^{(n)},\bm A^{(n)} \vert \phi_{\text{C}})$. The PSAC network acts as actor when the green area is activated, and critic when the pink area is activated, with the brown area being the shared part. The overall parameters are expressed as $\phi=\phi_{\text{A}}\bigcup \phi_{\text{C}}$. 
Note that the updating schemes across user and antenna dimensions differ slightly in the action encoder and action decoder due to customized hybrid action spaces. For brevity, we illustrate the update across the user dimension in Fig.~\ref{fig:network architecture}, while the update across the antenna dimension follows a similar pattern.

\begin{algorithm}[h]
  \caption{DRL-aided PSAC-based ISAC Precoding in double dynamics}
  \label{alg:ddpg}
  \begin{algorithmic}[1]
    \State Initialize PSAC network with random weights $\phi\!=\!\phi_{\text{A}}\!\bigcup \phi_{\text{C}}$; 
    \State Initialize the target network with $\phi' = \phi$;
    \State Initialize the experience replay buffer $\mathcal{M}$ with size 0, mini-batch size $N_b$, discount factor $\gamma$ and the adaptive variable with $\lambda=0.5$;
    \State Initialize a random process $\mathcal{N}$ for action exploration;
    \For {episode = 1 to $N_{\text{ep}}$}
    \State Initialize the observation $\bm o^{(1)}$;
    \For{$t=1$ to $T$}
      \State Select action $\bm a^{(n)}=\mu(\bm o^{(n)}\vert \phi_{\text{A}})$ with $\xi$-greedy policy, update $\bm F_{\text{RF}}$ and $\bm F_{\text{BB},m}$ based on $\bm a^{(n)}$; 
      \State Receive reward $r^{(n)}$ and observe the next state $\bm o^{(n+1)}$;
      \State Store the tuple $(\bm o^{(n)}, \bm a^{(n)},r^{(n)},\bm o^{(n+1)})$ into $\mathcal{M}$ and update $\mathcal{M}$;
      \EndFor
      \State Initialize the total critic loss as $L_{\text{c}}=0$;
      \For {$k=1$ to $K$}
      \State Sample a batch of $N_{\text{b}}$ tuples from $\mathcal{M}$ randomly;
      \State Compute $y_{i}$ and $\delta_{i}$ according to (\ref{equ:value_target}) and (\ref{equ:critic_loss});
      \State Update the PSAC network parameters through gradient descend according to (\ref{equ:update_network});
      \State Update critic loss as $L_{\text{c}}^{(k)}\!=\!L_{\text{c}}^{(k-1)}\!+\!\frac{1}{N_{\text{b}}}\sum_{i=1}^{N_{\text{b}}}\delta_{i}$;
      \State \textbf{if} $k$ mod $p$ \textbf{then}
       \State ~~~Average $L_{\text{c}}$, update $\lambda$ as (\ref{equ:lambda_update}) and reset $L_{\text{c}}\!=0$;
       \State ~~~\textbf{if} $\lambda>\lambda_{\text{thres}}$ \textbf{then}
        \State ~~~~~~Update the target network as $\phi'^{(k)}= \phi^{(k)}$;
    \EndFor
    \EndFor
  \end{algorithmic}
  \vspace{-0.1cm}
\end{algorithm}

\vspace{-0.2cm}
\subsection{Network Update}

The use of shared parameters in the PSAC architecture introduces an additional challenge in training stability compared to traditional actor-critic architectures. To address this challenge and enhance training stability, we have drawn inspiration from a previous study \cite{IAC} and implemented an adaptive loss function specifically designed for the PSAC network.
To reduce fluctuations in target values during training and expedite convergence, we employ target networks $Q'$ and $\mu'$ parameterized by $\phi_{\text{C}}'$ and $\phi_{A}'$ respectively. 
In the offline training stage, the transition tuple $(\mathcal{O}^{(n)}, \bm A^{(n)}, r^{(n)}, \mathcal{O}^{(n+1)})$ is stored in a memory replay buffer $\mathcal{M}$ for network updating.
During each episode, $K$ batches are sampled successively from $\mathcal{M}$ for network update.
When the $k$-th batch $(k=1,\cdots, K)$ of $N_{b}$ transition tuples, $\{(\bm o_{i}^{(k)}, \bm a_{i}^{(k)}, r_{i}^{(k)}, \bm o_{i+1}^{(k)} )\vert_{i=1}^{N_{\text{b}}} \} $, are randomly sampled, the target reward of the $i$-th sample ($i=1, \cdots, N_{b}$) is calculated as
\vspace{-0.2cm}
\begin{equation}
    y_{i}^{(k)}=r_{i}^{(k)}+\gamma Q'(\bm o^{(k)}_{i+1}, \mu'(\bm o^{(k)}_{i+1})).
    \label{equ:value_target}
    \vspace{-0.2cm}
\end{equation}
The critic loss and actor loss are calculated respectively as
\vspace{-0.2cm}
\begin{align}
    \delta_{\text{C},i}^{(k)}&=f(y_{i}^{(k)}-Q(\bm o_{i}^{(k)}, \bm a_{i}^{(k)})),
    \label{equ:critic_loss} \\
    \delta_{\text{A},i}^{(k)}&=f( \mu(\bm o^{(k)}_{i})-\!\mu'(\bm o^{(k)}_{i})), 
    \label{equ:actor_loss}
    \vspace{-0.2cm}
\end{align}
where $f(\cdot)$ is the $l$-1 smoothing function for enabling a more stable training, as it provides steady gradients for large value and hence helps to avoid exploding gradients.

Then we set the unified loss function for PSAC as $Z^{(k)}(\phi)=\frac{1}{N_{\text{b}}}\sum_{i=1}^{N_b}Z_{i}^{(k)}$, where
\vspace{-0.2cm}
\begin{equation}
    Z_{i}^{(k)}=\delta_{\text{C},i}^{(k)}+(1-\lambda^{(k)})\delta_{\text{A},i}^{(k)}-\lambda^{(k)} Q(\bm o^{(k)}_{i}, \mu(\bm o^{(k)}_{i})),
    \vspace{-0.2cm}
\end{equation}
including the critic loss, the actor loss along with the expected return based on the strategy.
$\lambda^{(k)} \in [0,1]$ is an adaptive variable to reflect the reliability of the critic, calculated as
\vspace{-0.2cm}
\begin{equation}
    \lambda^{(k)}=\tau e^{-{L^{(k)}_{\text{c}}}^{2}}+(1-\tau)\lambda^{(k-1)},
    \label{equ:lambda_update}
    \vspace{-0.1cm}
\end{equation}
where $L^{(k)}_{\text{c}}\!=\!\mathbb{E}_{\text{k},\text{i}}[\delta_{\text{C},i}^{(k)}]$ denotes the expected critic loss calculated in the $k$-th batch. $\tau \!\in\![0,1]$ is a hyper-parameter. Note that a larger $L_{\text{c}}^{(k)}$ results in a reduced $\lambda$, indicating a less reliable Q-value from the current critic. Within each batch, the PSAC networks are updated by minimizing the unified loss function through gradient descent as
\vspace{-0.1cm}
\begin{equation}
\begin{aligned}
    \phi^{(k)} &= \phi^{(k-1)} -\kappa \nabla_{\phi}Z^{(k)}(\phi) \\
    &=\phi^{(k-1)}-\frac{\kappa}{N_{b}}\sum_{i=1}^{N_{\text{b}}}\nabla_{\phi}Z^{(k)}_{i}(\phi),
    \label{equ:update_network}
\end{aligned}
\vspace{-0.2cm}
\end{equation}
with learning rate $\kappa$. $\lambda$ is updated every $p$ batches within one episode according to Eq.~(\ref{equ:lambda_update}), to re-evaluate the confidence of the critic.
Finally, the updating of the target networks as \cite{DDPG} is performed if $\lambda$ exceeds a certain threshold $\lambda_{\text{thres}}$.
The pseudo code of the whole training in detail is shown in Algorithm 2.

\vspace{-0.3cm}
\subsection{Convergence Analysis}

We then delve into the theoretical analysis of the convergence of the proposed scheme, to guarantee its stability. Note that PSAC is a specialized variant of the actor-critic architecture, with the actor sharing part of parameters with the critic. In fact, the shared parameters and objective functions do not affect the update of their respective networks. Therefore, the convergence analysis of the proposed PSAC is similar to that of the standard actor-critic \cite{on_AC, DDPG_convergence}.

\begin{proposition}
Denoting $J^{(t)}(\mu)$ as the expected cumulative reward achieved by the actor at the $t$-th iteration.
The output of the PSAC in Algorithm 2 satisfies
\vspace{-0.2cm}
\begin{equation}
\min_{t\in [1,T]} \mathbb{E}\|\nabla J^{(t)}(\mu)\|^{2}\leqslant \frac{c_{1}}{T} + \frac{c_{2}}{N_{\text{b}}}+c_{3}\kappa^{2},
\label{equ:pro_2}
\vspace{-0.2cm}
\end{equation}
where $c_{1}$, $c_{2}$, $c_{3}$ are constants and $\kappa$ denotes the approximation errors of the Q-value from the critic.
\end{proposition}
\vspace{-0.2cm}
\begin{proof}
    See Appendix B.
\end{proof}

Therefore, the policy gradient approaches zero as the number of episodes becomes sufficiently large.
The convergence upper bound of PSAC consists of the convergence rate term vanishing with iterations, the variance term caused by stochastic sampling and controllable by $N_{\text{b}}$, and the system error term composed of approximation errors. In practice, the high capacity of the neural networks can help to reduce the error and achieves better convergence accuracy.

\begin{table}[t]
    \centering
    \caption{System setup for environment generation}
    \begin{tabular}{p{1.6cm}p{2.0cm}||p{1.4cm} p{1.4cm}}
    \hline
       Parameter  &  value & Parameter  &  value \\
       \hline
       \hline
       $N_{t}$  & 32   &  $U^{\text{max}}$  & 16   \\
       \hline
       $N_{r}$ & 32   &  $G_{\text{t}}$  & 64   \\
       \hline
       $M$ &  32  & $N_{d}$  & 8   \\
       \hline
       $f_{c}$ &  28 GHz  &  $P_{u}$ &  8  \\
       \hline
       $\Delta f$ &  30 kHz  &  $\sigma_{\beta}$ &  1  \\
       \hline
       $L_{\text{cp}}$ &  8  & $\sigma_{\text{RCS}}$  & 10    \\
       \hline
       $L$ &  32  & $\sigma_{\text{c}}$  & -10 dBm   \\
       \hline
       $T$ &  100  &  $\sigma_{\text{s}}$ & -10 dBm    \\
       \hline
       $B$ &  3  & $d_{\text{max}}$  & 100 m   \\
       \hline
       $g_{\text{rp}}(\cdot)$ &  raised-cosine & $\vert \bm v \vert$ (m/s)  & $\mathcal{U}[10,30]$    \\
       \hline
    \end{tabular}
    \label{tab:environment}
\end{table}
\vspace{-0.1cm}

\begin{table}[t]
    \centering
    \caption{Hyper-parameters in model training}
    \begin{tabular}{p{2.9cm}||p{2.1cm}}
    \hline
       Parameter  &  value \\
       \hline
       \hline
       Discount factor $\gamma$  &  0.6 \\
       \hline
       Exploration rate $\xi$  &  0.2 \\
       \hline
       Batch size $N_{\text{b}}$  &  64 \\
       \hline
       Learning rate $\kappa$  &  0.02 \\
       \hline
       Replay buffer size  &  1024 \\
       \hline
       Optimizer  &  Adam \\
       \hline
       Batches per episode $K$  &  16 \\
       \hline
       Updating period $p$  &  4 \\
       \hline
       Initial critic reliability $\lambda$  &  0.5 \\
       \hline
       Update weight $\tau$  &  0.5 \\
       \hline
       Precoding update step $\eta_{r}$  &  0.1 \\
       \hline
    \end{tabular}
    \label{tab:model}
    \vspace{-0.4cm}
\end{table}

\section{Simulation Results}
\vspace{-0.1cm}
In this section, experimental results are presented to evaluate the performance of the proposed ISAC precoding schemes in doubly-dynamic scenarios.

We first construct a joint GPS-CSI dataset to simulate doubly dynamic vehicular network scenarios, which serves as the platform for training and testing the algorithms.
\begin{itemize}
    \item \textbf{Creation of the scenario}: We utilize the Simulation of Urban MObility (SUMO) software to simulate the dynamic trajectories of both moving vehicles and the low-altitude UAV at an intersection \cite{SUMO}, where the road width is 30 m. The BS covers a lane length of 180 meters horizontally and 80 meters vertically. The vehicle trajectories include straight motion, turning and lane-changing, while the UAV trajectories encompass horizontal straight-line flight, vertical lift and hovering.
    \item \textbf{Construction of prior positioning information}: Once the scene is constructed, the coordinates of the vehicles and the UAV within the intersection are exported. The GPS feedbacks from the users are generated by adding noise to their coordinates, simulating GPS errors in real-world scenarios.
    \item \textbf{Generation of channel data}: After importing the scene into Wireless InSite® software \cite{WI}, communication and sensing channels for each symbol duration are generated using ray-tracing, based on the channel model outlined in Section II. This channel information serves as the foundation for calculating the loss and training the model.
\end{itemize}

\vspace{-0.1cm}
The system setup for environment generation and the hyper-parameters setup for proposed DRL-aided scheme in simulations can be found in Table~\ref{tab:environment} and Table~\ref{tab:model}, respectively. 
The methods to be evaluated includes:
\vspace{-0.1cm}
\begin{itemize}
    \item \textbf{DRL-PC}: The proposed scheme with both position-related and channel-related observations as the  input.
    \item \textbf{DRL-PO}: The proposed scheme with only position-related observation as the input.
    \item \textbf{DRL-CO}: The proposed scheme with only channel-related observation as the input.
    \item \textbf{DRL-UU}: the proposed scheme with User-dimension Updating as the action.
    \item \textbf{DRL-AU}: The proposed scheme with Antenna-dimension Updating as the action.
    \item \textbf{Opt-based}: The optimization-based wideband precoding with perfect state information proposed in Sec. III.
    \item \textbf{MP-based}: Predictive precoding scheme in \cite{dynamic_MP}.
    \item \textbf{Greedy-based}: Precoding scheme proposed in \cite{shijian_MI}.
\end{itemize}

\begin{figure}[t]
  \vspace{-0.2cm}
  \centering
   \subfigure[Reward of different schemes.\label{fig:reward_1}]
 {\includegraphics[width=0.66\linewidth]{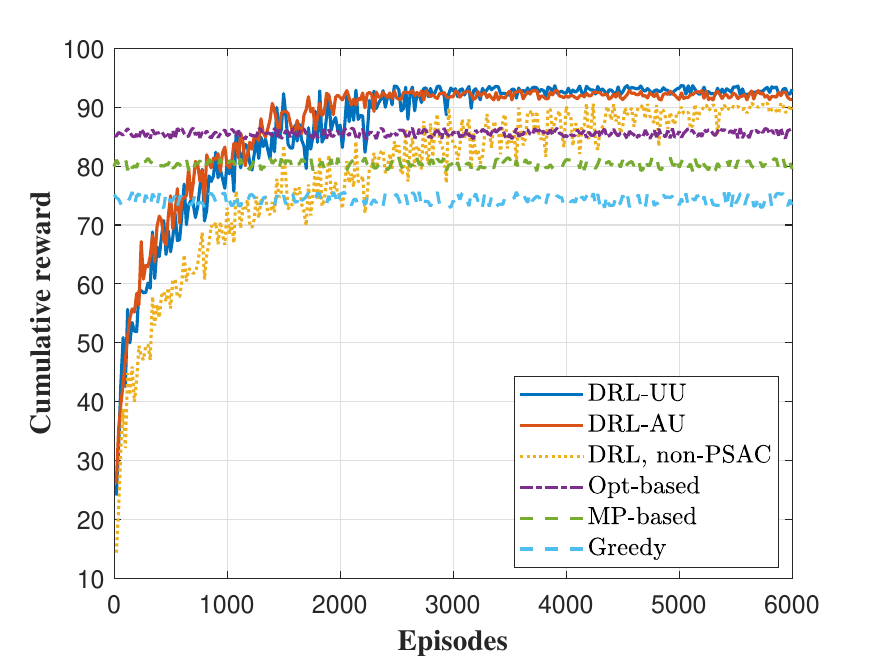}}
  \subfigure[Reward from different inputs.\label{fig:reward_2}]
 {\includegraphics[width=0.66\linewidth]{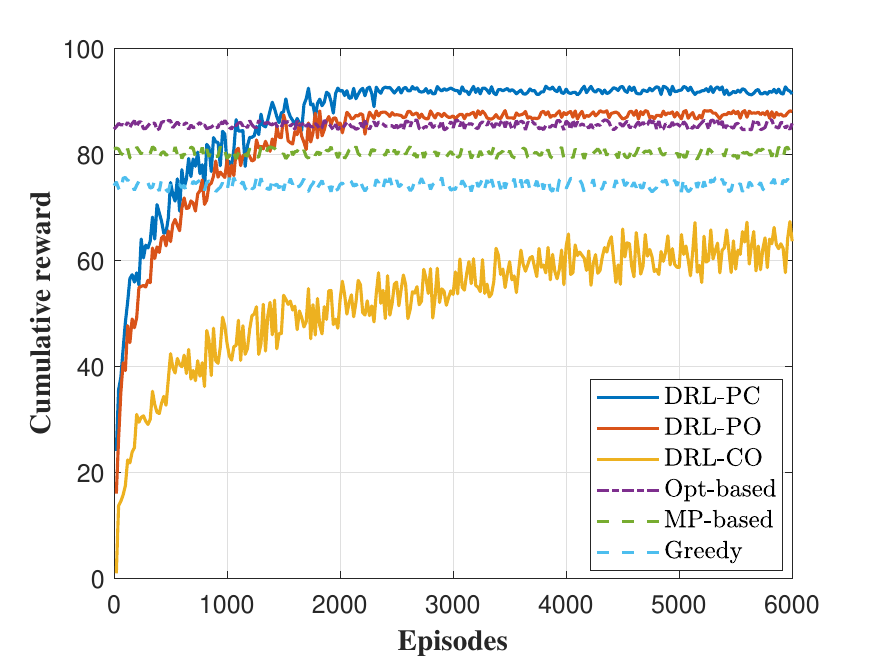}}
    \vspace{-0.2cm}
    \captionsetup{font={small}}
  \caption{Cumulative reward comparison among different schemes.}
  \label{fig:reward}
  \vspace{-0.5cm}
\end{figure}

The proposed algorithm is developed using PyTorch and implemented on a Window™-based machine equipped with an NVIDIA™ GeForce RTX4060 GPU for efficient processing.

\vspace{-0.2cm}
\subsection{Cumulative Reward Versus Episodes in the Training}

Firstly, we verify BS’s achievable cumulative reward within one frame in the training process of the proposed schemes in Fig.~6. We set SNR as 0 dB and $U=8$.
As the number of episodes increases, the cumulative reward of the proposed algorithm gradually rises. After convergence, the cumulative reward of the DRL-PC can surpass the optimization-based method with perfect state information and MP-based benchmarks. This reflects that the proposed DRL-aided method can alleviate the performance loss from the relaxation in optimization-based algorithms and  mitigate the reliance on instantaneous state information in double dynamics.
Even when relying solely on historical position estimates in DRL-PO version, the proposed scheme still achieves high reward, keeping a narrow gap between DRL-PC. However, merely initial CSI as input in DRL-CO is not enough for BS in the training, leading to degraded and unstable cumulative reward.

Compared to the training of the classic actor-critic frameworks, the adopted PSAC architecture exhibits much faster convergence speed, illustrating its advantage in accelerating training in complicated state space and action space.
Additionally, DRL-AU scheme exhibits a faster convergence compared to DRL-UU scheme, reflecting its higher updating efficiency. Therefore, the DRL-UU is better suited for environments with more stable conditions, while the DRL-AU is more appropriate for dynamic scenarios involving rapid changes in the distribution of targets and users.

We also compare the impact of hyper-parameters on the convergence speed and the achievable rewards in the training of DRL-AU in Table.~\ref{tab:discount} and Table.~\ref{tab:exploration}. Both excessively high and low discount factors $\gamma$ lead to reduced performance after convergence. A higher $\xi$ results in a slower convergence, while a lower $\xi$ accelerates convergence but causes the model to reside at a local optimum, thereby reducing the achievable cumulative reward. Therefore, appropriately selecting hyper-parameters is crucial for improving both model training efficiency and performance.

\begin{table}
    \centering
    \caption{Impact of $\gamma$ on the training}
    \vspace{-0.2cm}
    \begin{tabular}{p{2.2cm}||p{0.45cm}||p{0.45cm}||p{0.45cm}||p{0.45cm}||p{0.45cm}||p{0.45cm}}
    \hline
       $\gamma$  &  0 &  0.2 &  0.4 &  0.6 &  0.8 &  1.0 \\
       \hline
       \hline
       Converging epochs & 1810 &  1840 & 1870 & 1880 &  1880 & 1910 \\
       \hline
       Cumulative reward &  84.69 &  88.52 &  90.77 &  91.56 &  90.34 &  87.21 \\
       \hline
    \end{tabular}
    \label{tab:discount}
    \vspace{-0.1cm}
\end{table}

\begin{table}
    \centering
    \caption{Impact of $\xi$ on the training}
    \vspace{-0.2cm}
    \begin{tabular}{p{2.2cm}||p{0.45cm}||p{0.45cm}||p{0.45cm}||p{0.45cm}||p{0.45cm}||p{0.45cm}}
    \hline
       $\xi$  &  0 &  0.2 &  0.4 &  0.6 &  0.8 &  1.0 \\
       \hline
       \hline
       Converging epochs & 1710 &  1880 & 2420 & 3510 & 4180 &  - \\
       \hline
       Cumulative reward &  86.04 &  91.56 &  91.57 &  91.57 &  91.57 &  - \\
       \hline
    \end{tabular}
    \label{tab:exploration}
    \vspace{-0.4cm}
\end{table}

\begin{figure}[t]
  \vspace{-0.65cm}
  \centering
   \subfigure[Averaged CRLB of different schemes.\label{fig:CRB_scheme}]
    {\includegraphics[width=0.66\linewidth]{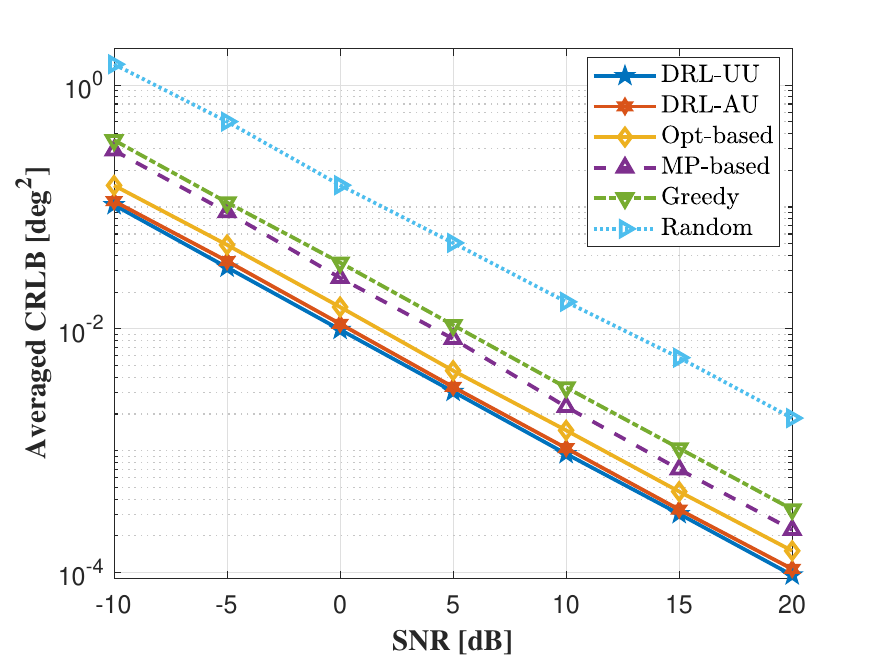}}
  \subfigure[Averaged SE of different schemes.\label{fig:SE_scheme}]
    {\includegraphics[width=0.66\linewidth]{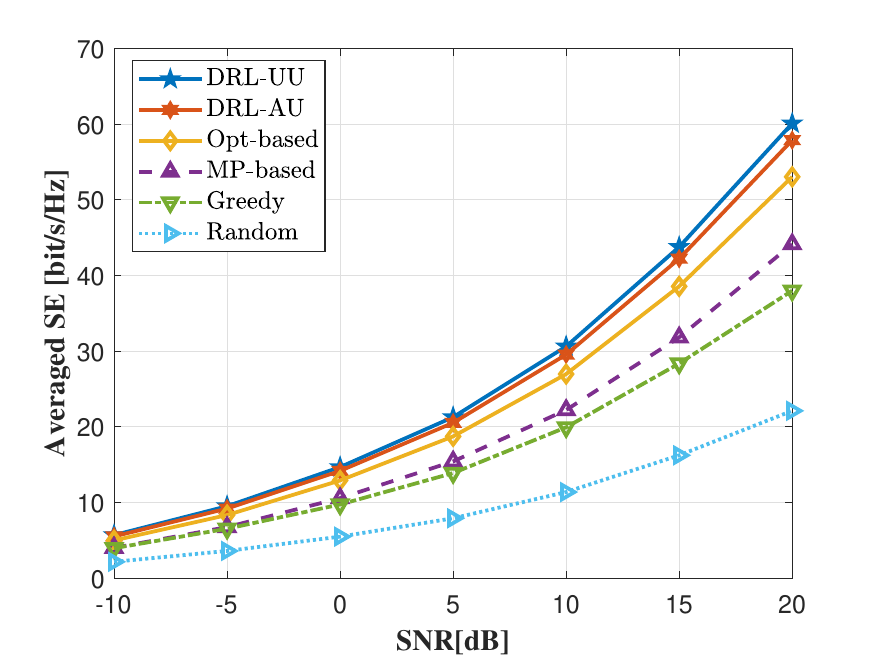}}
    \vspace{-0.2cm}
    \captionsetup{font={small}}
  \caption{ISAC performance comparison among different schemes.}
  \label{fig:reward}
  \vspace{-0.25cm}
\end{figure}

\begin{figure*}[t]
\vspace{-0.0cm}
\centering
\captionsetup{font={small}}
\begin{minipage}[b]{.3\linewidth}
\includegraphics[scale=0.39]{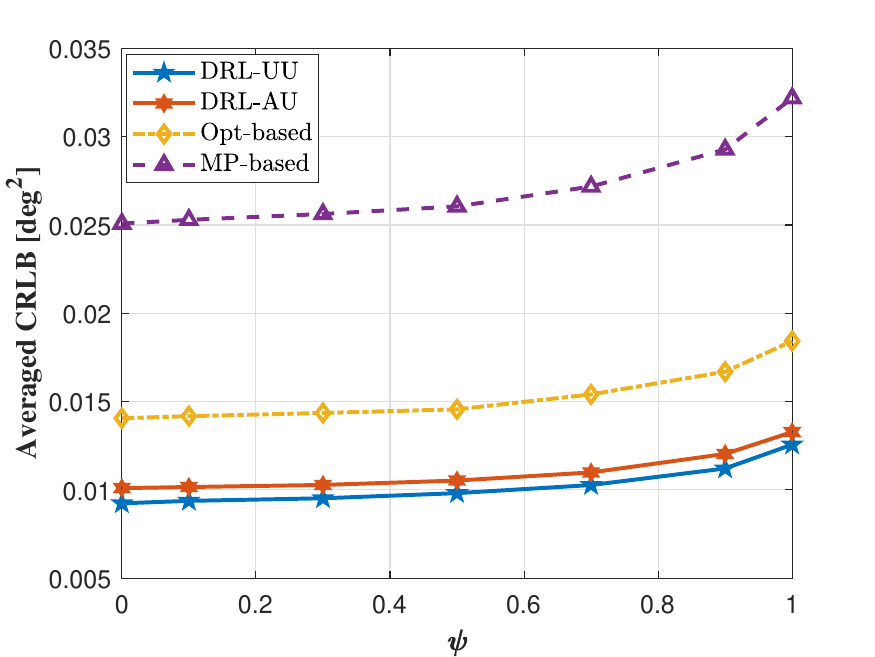}
\caption{Averaged CRLB versus $\psi$.}
\label{fig:crb_psi}
\end{minipage}
\begin{minipage}[b]{.3\linewidth}
\includegraphics[scale=0.39]{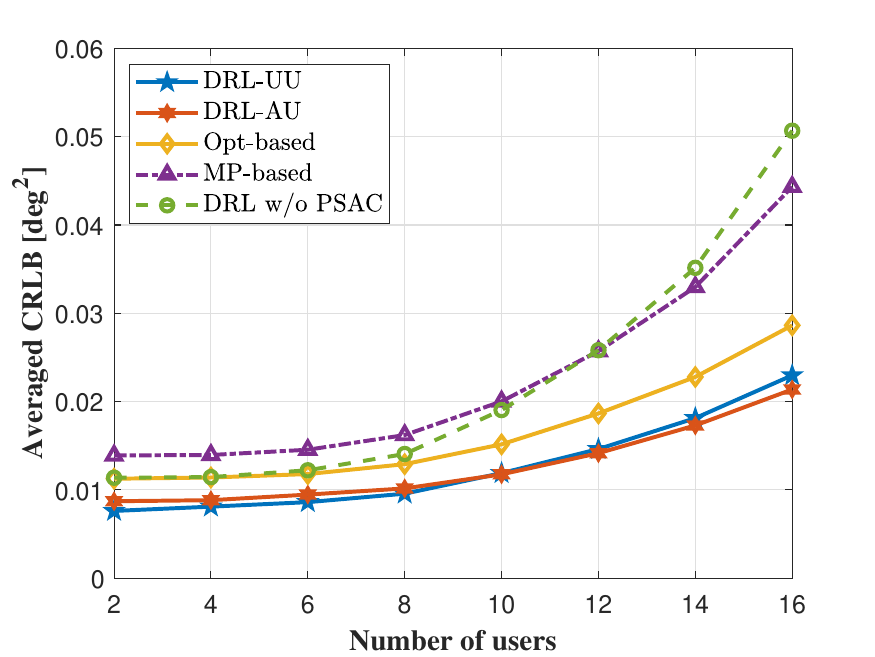}
\caption{Averaged CRLB versus $U$.}
\label{fig:crb_u}
\end{minipage}
\begin{minipage}[b]{.3\linewidth}
\includegraphics[scale=0.39]{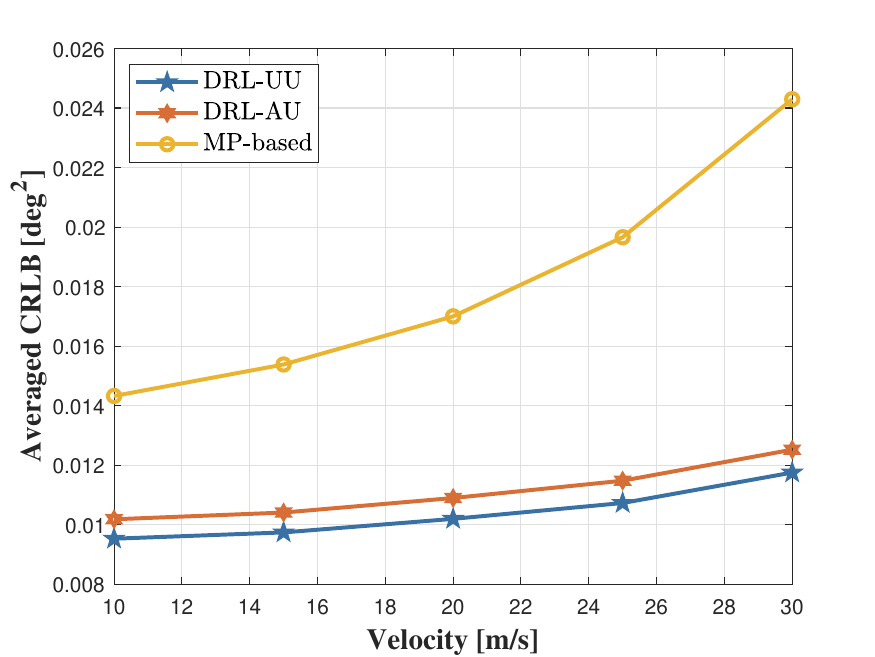}
\caption{Averaged CRLB versus velocity.}
\label{fig:crb_v}
\end{minipage}
\begin{minipage}[b]{.3\linewidth}
\includegraphics[scale=0.39]{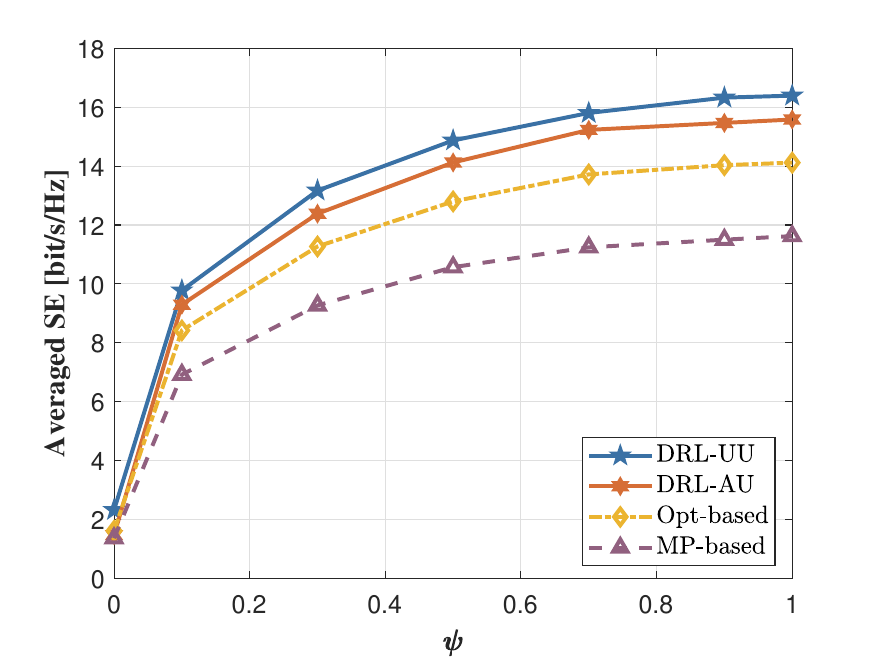}
\caption{Averaged SE versus $\psi$.}
\label{fig:SE_psi}
\end{minipage}
\begin{minipage}[b]{.3\linewidth}
\includegraphics[scale=0.39]{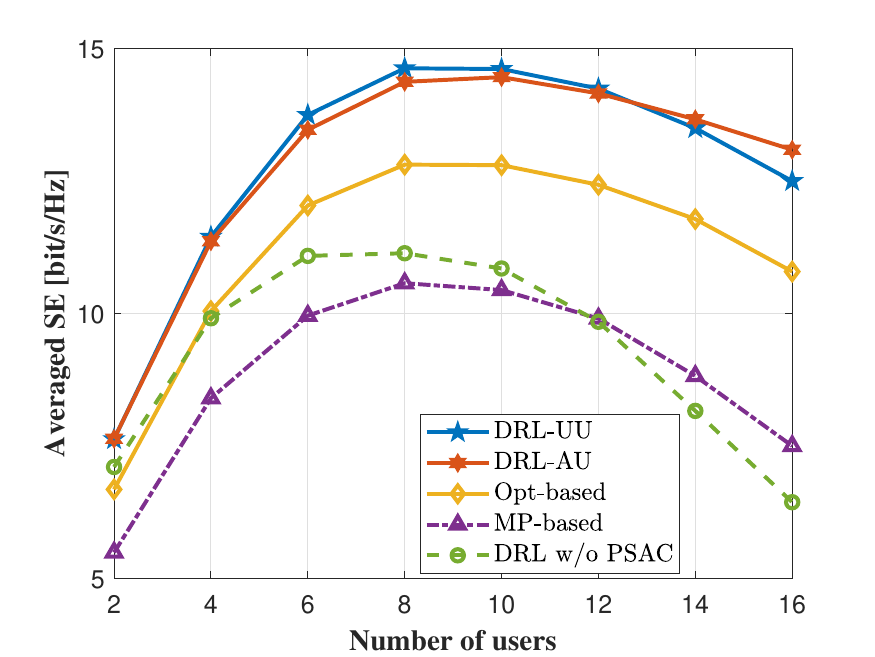}
\caption{Averaged SE versus $U$.}
\label{fig:SE_u}
\end{minipage}
\begin{minipage}[b]{.3\linewidth}
\includegraphics[scale=0.39]{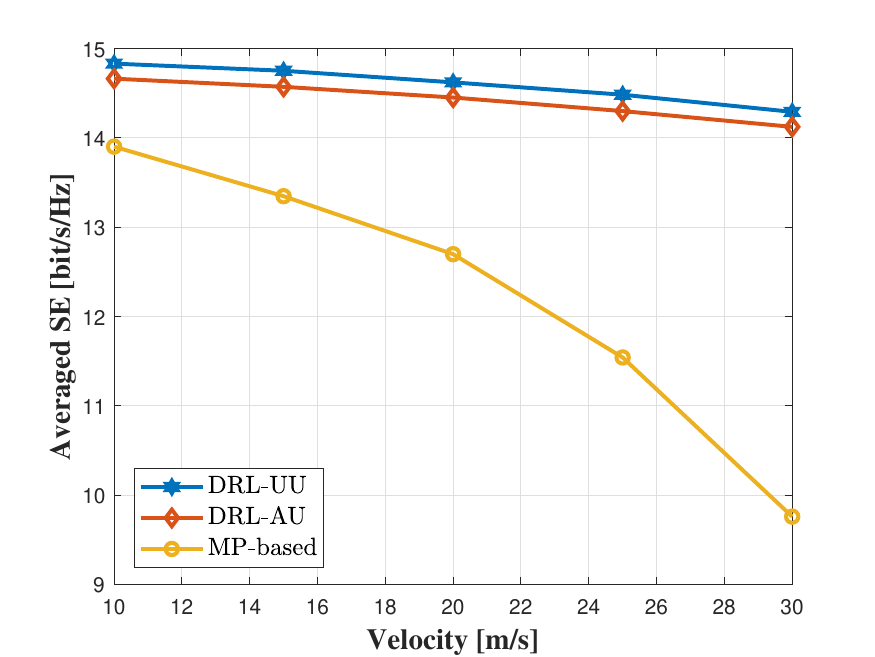}
\caption{Averaged SE versus velocity.}
\label{fig:SE_v}
\end{minipage}
\vspace{-0.1cm}
\end{figure*}

\vspace{-0.2cm}
\subsection{ISAC Performance in Different Schemes}
We then verify the ISAC performance of the proposed schemes versus SNR after convergence in Fig.~\ref{fig:CRB_scheme} and Fig.~\ref{fig:SE_scheme}. In terms of target tracking, DRL-aided schemes may implicitly predict the current target position based on existing estimates, thereby reducing the averaged CRLB of the target's angle estimation.
The DRL-aided schemes yield a performance gain of 2 dB compared to Opt-based, considerably improving the sensing accuracy.

With respect to communications, we compare the averaged SE of different schemes. Despite lacking the instantaneous CSI at each subframe, the BS can adjust the precoding matrix through interaction with the environment to gradually keep up with the changes of the channels.
The proposed algorithm consistently outperforms the Opt-based method by a performance gain of almost 2.5 dB.

Moreover, the ISAC performance of the DRL-aided scheme closely approaches the ideal case when perfect channel prior information is used as input rather than the initial CSI. This suggests that the proposed scheme does not purely rely on perfect prior information. Additionally, the performance in this case surpasses that of the Opt-based approach, providing a more equitable comparison that highlights the superiority of the DRL-aided scheme.

\vspace{-0.2cm}
\subsection{ISAC Performance Versus the Weighting Coefficient}

We proceed to examine the influence of the weighting coefficient $\psi$ on the ISAC performance after convergence in Fig.~\ref{fig:crb_psi} and Fig.~\ref{fig:SE_psi}, fixing SNR= 0 dB. A larger $\psi$ indicates a higher priority given to communication performance. As $\psi$ increases, BS tends to prioritize boosting the SE to attain higher rewards during training. This leads to an enhancement in the averaged SE at the cost of a slight deterioration in CRLB.
Conversely, when $\psi$ is smaller, BS prioritizes sensing performance, which may result in a slight reduction in the SE but an improvement in CRLB. Therefore, the flexible adjustment of $\psi$ enables the fulfillment of various ISAC performance indicators in real-world scenarios.

\vspace{-0.3cm}
\subsection{ISAC Performance Versus the Number of Users}

Furthermore, we investigate the impact of the number of users on ISAC performance in Fig.~\ref{fig:crb_u} and Fig.~\ref{fig:SE_u}, while keeping SNR at 0 dB, $\psi$ at 0.5. In terms of sensing performance, an increased $U$ introduces a more intricate trade-off between multi-user communications and target sensing. Consequently, the BS may compromise the accuracy of sensing to maintain the QoS for multi-user communications, resulting in an increased averaged CRLB.

Regarding communication performance, the average SE initially rises as the number of users increases. However, when the number of users surpasses a certain threshold, the difficulty of IUI elimination amplifies, resulting in a decline in the average SE. The proposed algorithm maintains its superiority over the benchmark algorithms across varying user counts.
Thanks to the dedicated design of the hybrid action space for hybrid precoding combined with the PSAC architecture,  
the proposed DRL-aided algorithm exhibits greater adaptability to changes in the number of users, preventing the occurrence of the crash in the SE observed in non-PSAC architectures (DQN in the experiments) when the number of users becomes excessively large. 
Furthermore, the DRL-AU scheme demonstrates less sensitivity to an increase in the number of users compared to the DRL-UU scheme. This difference can be attributed to the action space design of the DRL-AU scheme, which remains independent of the number of users.

\vspace{-0.3cm}
\subsection{Impact of Environmental Double Dynamics}

Additionally, we assess the robustness of the proposed schemes under double dynamics in Fig.~\ref{fig:crb_v} and Fig.~\ref{fig:SE_v}. We conduct simulations of the ISAC performance while maintaining a fixed velocity ranging from 10 m/s to 30 m/s. In the MP-based tracking scheme, the ISAC performance shows a significant decline as the velocity increases. This decrease can be attributed to the increased difficulty of tracking time-varying multi-user channels and targets under high dynamics.

In contrast, the ISAC performance of the DRL-aided scheme demonstrates only a slight decrease at higher speeds, highlighting the potential of DRL-aided methods in effectively adapting to the challenges posed by double dynamics. This resilience to high-speed scenarios showcases the strength of DRL-based approaches in handling dynamic and complex environments.

\vspace{-0.2cm}
\subsection{Performance Boundary Comparison}
\begin{figure}[t]
  \vspace{0cm}
  \centering
  \includegraphics[width=0.69\linewidth]{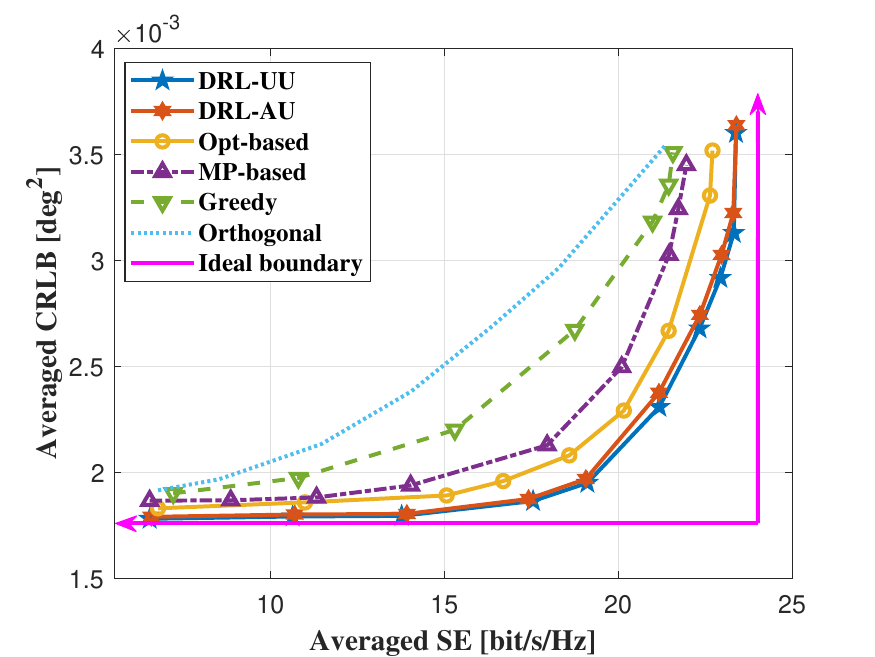}
  \captionsetup{font={small}}
  \caption{ISAC performance boundary of different methods.}
  \label{fig:bound}
  \vspace{-0.3cm}
\end{figure}

As a step further, we describe the achievable ISAC performance boundary in Fig.~\ref{fig:bound}, by changing the weighting coefficient in the reward function and connecting the achievable averaged SE-CRLB points into a curve.
The proposed scheme achieves a significant gain compared to the existing benchmarks in terms of trade-off, approaching the ideal boundary. Compared to that under the completely orthogonal operations where the communication and sensing are separate, the achieved bound is considerably enhanced, reflecting the cooperation gain of ISAC, as analyzed in \cite{JCR_ISIT}.

\vspace{-0.3cm}
\subsection{Complexity Analysis and Comparison}

The computational complexity of different schemes are compared in Table.~\ref{tab:complexity}.
The required number of floating point operations (FLOPs) is used as the metric for complexity.  
In the online deployment of the proposed DRL-aided scheme, only the trained actor is activated for updating $\bm F_{\text{RF}}$ and $\bm F_{\text{BB}}$ via forward pass. Denote $L_{\text{c}}$ as the maximal number of convolutional layers, $N_{\text{f}}$ as the maximal number of input and output feature maps, and $N_{\text{k}}$ as the maximal side length of the filters in convolutional layers.
$N_{\text{W}}$ and $N_{\text{L}}$ as the maximal number of neurons in the hidden layers and the total number of hidden layers in fully-connected layers. 
Then the complexity in the forward pass of the actor network is $\mathcal{O}(L_{\text{c}}(UG_{\text{t}}N_{\text{d}}+N_{\text{x}}N_{\text{y}})N_{\text{k}}^{2}N_{\text{f}}^{2} +N_{\text{L}}^{2}N_{\text{W}})$, which is linear to $U$ and independent with $N_{\text{t}}$, $M$. Based on the action, the precoding is updated with the complexity $\mathcal{O}(N_t+M)$ for DRL-UU and $\mathcal{O}(UM)$ for DRL-AU. Hence, the DRL-UU demonstrates lower complexity in systems with a large number of users, whereas the DRL-AU exhibits lower complexity in systems with large antenna arrays.

\begin{figure}[t]
  \vspace{0cm}
  \centering
  \includegraphics[width=0.69\linewidth]{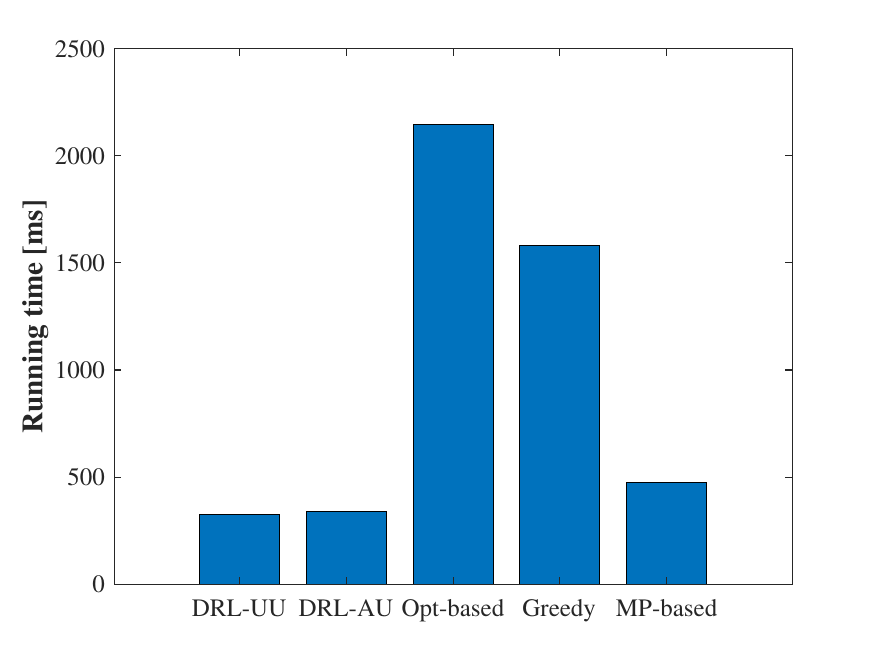}
  \captionsetup{font={small}}
  \caption{Running time comparisons.}
  \label{fig:time}
  \vspace{-0.3cm}
\end{figure}

We further compared the averaged running times within one frame of different schemes in Fig.~\ref{fig:time}. The proposed DRL-aided algorithm demonstrates minimal complexity and considerably shorter running times compared to other benchmarks, highlighting its advantage in reducing complexity.
Additionally, the DRL-AU scheme boasts a shorter runtime in comparison to the DRL-UU scheme.

\begin{table}[h]
\vspace{-0.1cm}
    \centering
    \caption{Complexity comparison}
    \begin{tabular}{|p{2.2cm}|p{5.7cm}|}
    \hline
       Method  &  Complexity \\
       \hline
       \hline
       DRL-UU  &  $\mathcal{O}(L_{\text{c}}(UG_{\text{t}}N_{\text{d}}\!+\!N_{\text{x}}N_{\text{y}})\!N_{\text{k}}^{2}N_{\text{f}}^{2} \!+\!N_{\text{w}}^{2}N_{\text{L}}+N_{t}+M)$       \\
       \hline
       DRL-AU  &  $\mathcal{O}(L_{\text{c}}(UG_{\text{t}}N_{\text{d}}\!+\!N_{\text{x}}N_{\text{y}})\!N_{\text{k}}^{2}N_{\text{f}}^{2} \!+\!N_{\text{w}}^{2}N_{\text{L}}+UM)$       \\
       \hline
       Opt-based  &  $\mathcal{O}(M(N_{t}^{3.5}\log(1/\epsilon)+2N_{\text{iter}}U^{2}N_{t}))$     \\
       \hline
       MP-based  &  $\mathcal{O}(4UN_{\text{iter}}+U^{2}N_{t})$     \\
       \hline
       Greedy-based &  $\mathcal{O}(M(U^{4}+U^{2})N_{\text{t}})$      \\
       \hline
    \end{tabular}
    \label{tab:complexity}
    \vspace{-0.3cm}
\end{table}

\vspace{-0.1cm}

\section{Conclusion}

This paper investigated the ISAC hybrid precoding for scenarios with time-varying communication and sensing channels. We proposed an optimization-based method for wideband multi-user ISAC systems with full environmental knowledge and a SoM-enhanced DRL-based precoding approach for the cases with imperfect state information. By leveraging different modalities, including CSI and GPS, our method adapts to dynamic changes effectively. The enhanced training algorithm is able to navigate complex action spaces and adjust to varying user numbers. The devised hybrid precoding update scheme spans both the user and the antenna dimensions, leading to improved ISAC performance in doubly-dynamic scenarios while reducing reliance on real-time information and computational burdens. The multifaceted advantages were validated by extensive simulations.

\vspace{-0.1cm}

\appendices
\section{Proof of the proposition 1}

Denote $\widehat{\bm W}_{m,u}$ as the global optimum to the relaxed version from (\ref{Problem4}). The precoding vectors are constructed as
\begin{equation}
    \widetilde{\bm f}_{m,u}=(\bm h_{m,m,u}^{\text{H}}\widehat{\bm W}_{m,u}\bm h_{m,m,u})^{-\frac{1}{2}}\widehat{\bm W}_{m,u}\bm h_{m,m,u},
\end{equation}
for $m\in [1,M]$ and $u\in [1,U]$. Then the covariance are $\widetilde{\bm W}_{m,u}=\widetilde{\bm f}_{m,u}\widetilde{\bm f}_{m,u}^{\text{H}}$. It is obvious that $\widetilde{\bm W}_{m,u}$ is positive semi-definite and rank-1. Since the objective function in (\ref{Problem4}) is not directly influenced by $\widetilde{\bm W}_{m,u}$, we only need to verify that $\widetilde{\bm W}_{m,u}$ is a feasible solution to (\ref{Problem4}).

Firstly, it can be derived that 
\vspace{-0.2cm}
\begin{align}
    \bm h^{\text{H}}_{m,m,u}\widetilde{\bm W}_{m,u}\bm h_{m,m,u}
    =&\bm h^{\text{H}}_{m,m,u}\widetilde{\bm f}_{m,u}\widetilde{\bm f}^{\text{H}}_{m,u}\bm h_{m,m,u} \nonumber \\
    =&\bm h^{\text{H}}_{m,m,u}\widehat{\bm W}_{m,u}\bm h_{m,m,u},
\end{align}
while $\bm h^{\text{H}}_{m,k,u}\widetilde{\bm W}_{k,i}\bm h_{m,k,u} \leqslant \bm h^{\text{H}}_{m,k,i}\widetilde{\bm W}_{k,i}\bm h_{m,k,i}$ when $k\neq m$ or $i\neq u$.
Thus $\text{Tr}(\bm Q_{m,k,u}\widetilde{\bm W}_{k,i})\leq\text{Tr}(\bm Q_{m,m,u}\widehat{\bm W}_{m,u})$, taking equal if and only if $k=m$ and $i=u$. Substituting these to (\ref{Problem4_d}) gives rise to
\vspace{-0.2cm}
\begin{align}
    &\text{Tr}(\bm Q_{\text{m,m,u}}\!\widetilde{\bm W}_{\text{m,u}})\!\!-\!\tau\!\!\! \sum_{i=1,i\neq u}^{U}\!\!\text{Tr}(\bm Q_{\text{m,m,u}}\widetilde{\bm W}_{\text{m,i}})
    \!\!-\!\tau \!\!\!\sum_{k=1, k\neq m}\!\!\!\!\bm Q_{\text{m,k,u}}\widetilde{\bm W}_{\text{k,u}} \nonumber \\
    &\geqslant \!\!\text{Tr}(\!\bm Q_{\text{m,m,u}}\!\widehat{\bm W}_{\text{m,u}})\!\!-\!\!\tau\!\!\! \sum_{i=1,i\neq u}^{U}\!\!\!\text{Tr}(\bm Q_{\text{m,m,u}}\!\widehat{\bm W}_{\text{m,i}})\!\! -\!\tau\!\! \!\sum_{k=1, k\neq m}\!\!\!\!\bm Q_{\text{m,k,u}}\!\widehat{\bm W}_{\text{k,u}} \nonumber \\
    &\geqslant \tau \sigma_{\text{c}}^{2},
    \vspace{-0.2cm}
\end{align}
namely, SE-related constraint (\ref{Problem4_d}) holds for $\widetilde{\bm W}_{m,u}$.

Subsequently, we prove that the CRB-related constraint also holds for $\widetilde{\bm W}_{m,u}$.
Notice that $\text{Tr}(\dot{\bm A}^{\text{H}}\dot{\bm A}\widetilde{\bm W}_{m,u})=\text{Tr}(\dot{\bm A}^{\text{H}}\dot{\bm A}(\bm h_{m,m,u}^{\text{H}}\widehat{\bm W}_{m,u}\bm h_{m,m,u})^{-1}\widehat{\bm W}_{m,u}\bm h_{m,m,u}\bm h_{m,m,u}^{\text{H}}\widehat{\bm W}_{m,u})=\text{Tr}(\dot{\bm A}^{\text{H}}\dot{\bm A}\widehat{\bm W}_{m,u})$, it can be derived that
\vspace{-0.2cm}
\begin{align}
&\left[\text{Tr}(\dot{\bm A}^{\text{H}}\dot{\bm A}\widetilde{\bm W}_{m,u})\!-\!\zeta \right]\text{Tr}(\bm A^{\text{H}}\bm A\widetilde{\bm W}_{m,u})\!-\!\vert\text{Tr}(\dot{\bm A}^{\text{H}}\bm A\widetilde{\bm W}_{m,u}) \vert^{2} \nonumber\\
=&\left[\text{Tr}(\dot{\bm A}^{\text{H}}\dot{\bm A}\widehat{\bm W}_{m,u})-\zeta \right]\text{Tr}(\bm A^{\text{H}}\bm A\widetilde{\bm W}_{m,u}) \geqslant 0,
\label{equ:crb_derive}
\end{align}
given the orthogonality between column vectors in $\bm A$ and $\dot{\bm A}$. By summing (\ref{equ:crb_derive}) over $u$, it can be proved that the constraint (\ref{Problem4_c}) holds.

Finally, we will examine the power constraint as follows:
\vspace{-0.2cm}
\begin{align}
    &\sum_{m=1}^{M}\text{Tr}(\tilde{\bm W}_{m})=\sum_{m=1}^{M}\sum_{u=1}^{U}\text{Tr}(\tilde{\bm W}_{m,u}) \nonumber \\
    =&\!\!\!\sum_{m=1}^{M}\!\!\sum_{u=1}^{U}\!(\bm h_{m,m,u}^{\text{H}}\!\!\hat{\bm W}_{m,u}\bm h_{m,m,u}\!)^{-1}\text{Tr}(\bm h_{m,m,u}^{\text{H}}\!\!\hat{\bm W}_{m,u}^{\text{H}}\!\!\hat{\bm W}_{m,u}\bm h_{m,m,u}\!) \nonumber \\
    \leqslant &\!\!\! \sum_{m=1}^{M}\!\sum_{u=1}^{U} \!\text{Tr}(\bm h_{m,m,u}\bm h_{m,m,u}^{\text{H}}\!\!\hat{\bm W}_{m,u})^{-1} \!\text{Tr}(\!\bm h_{m,m,u}\bm h_{m,m,u}^{\text{H}}\!\!\hat{\bm W}_{m,u})\nonumber \\
    &\text{Tr}(\hat{\bm W}_{m,u}) =\text{Tr}(\hat{\bm W}_{m,u})\leqslant P_{\text{t}},
    \vspace{-0.2cm}
\end{align}
according to the operator norm inequality. Hence the power constraint (\ref{Problem4_e}) also holds.

With the derivation above, it is verified that $\tilde{\bm W}_{m,u}$ is a feasible solution, and furthermore, it is also a global optimum to (\ref{Problem4}), thereby completing the proof.

\section{Proof of the Proposition 2}
Under the assumptions in \cite{DDPG_convergence}, the policy gradient $\nabla J(\mu)$ is Lipschitz continuous, laying the foundation for the subsequent finite-sample analysis.

First, we characterize the propagation of the dynamics of critic’s tracking error by coupling it with the actor’s updates.
Using the time-difference learning property, the update of the critic network's parameter $\phi_{\text{C}}$ follows
\begin{equation}
    \mathbb{E}\|\phi_{\text{C}}^{(t+1)}\!-\! \phi_{\text{C}\vert \mu}^{(t)\star} \|^{2}\!\leqslant\! (1\!-\!\alpha_{1}\kappa_{\text{C}})\mathbb{E}\|\phi_{\text{C}}^{(t)}\!- \!\phi_{\text{C}\vert \mu}^{(t)\star} \|^{2}+\!\frac{\alpha_{2}\kappa_{\text{C}}^{2}}{N_{\text{b}}},
\end{equation}
with $\alpha_{1}$, $\alpha_{2}$ being constants. Taking into account the actor's update renders
\vspace{-0.2cm}
\begin{equation}
\begin{aligned}
    \mathbb{E}&\|\phi_{\text{C}}^{(t+1)}- \phi_{\text{C}\vert \mu}^{(t+1)\star} \|^{2}\leqslant  \frac{\alpha_{3}}{\kappa_{\text{C}}}\mathbb{E}\|\phi_{\text{A}}^{(t+1)}-\phi_{\text{A}}^{(t)}\|^{2} \\
    &+ (1-\frac{\alpha_{1}\kappa_{\text{C}}}{2})\mathbb{E}\|\phi_{\text{C}}^{(t)}- \phi_{\text{C}\vert \mu}^{(t+1)\star} \|^{2}+\frac{2\alpha_{2}\kappa_{\text{C}}^{2}}{N_{\text{b}}},
\end{aligned}
\end{equation}
where $\phi_{\text{C} \vert \mu}^{(t) \star}$ is the global optimum of TD for a given fixed policy for a given $\mu$, and $\alpha_{3}$ is another constant.

Subsequently, the cumulative tracking error can be bounded by considering the dynamics of the tracking error. From the update rule of the actor parameter $\phi_{\text{A}}$, we have
\vspace{-0.2cm}
\begin{align}
    \phi_{\text{A}}^{(t+1)}-\phi_{\text{A}}^{(t)}&=\frac{\kappa_{\text{A}}}{N_{\text{b}}}\sum_{n=1}^{N_{\text{b}}}\nabla_{\phi_{\text{A}}}\mu^{(t)}(\bm o_{t,n})\nabla_{\phi_{\text{A}}}\mu^{(t)}(\bm o_{t,n})^{\text{T}}\phi_{\text{C}}^{(t)} \nonumber \\
    &:=\alpha_{\mu}h_{\mu}(\phi_{\text{C}}^{(t)}).
\end{align}
By exploiting the property of the Fisher information of the defined policy, the estimation bias is characterized as
\begin{equation}
\vspace{-0.2cm}
    \mathbb{E}\|h_{\mu}(\phi^{(t)}_{\text{C}})\!-\!\nabla J^{(t)}(\mu) \|^{2}\!\leqslant \!\alpha_{4}\mathbb{E}\|\phi_{\text{C}}^{(t)}\!- \!\phi_{\text{C}\vert \mu}^{(t)\star} \|^{2}\!+\!\alpha_{4}\kappa^{2}\!+\!\frac{\alpha_{5}}{N_{\text{b}}},
\end{equation}
with $\alpha_{4}$, $\alpha_{5}$ being constants. Then the cumulative dynamic tracking error is expressed as
\vspace{-0.2cm}
\begin{align}
\sum_{t = 0}^{T - 1}& \mathbb{E} \left\lVert \phi_{\text{C}}^{(t)}- \phi_{\text{C}\vert \mu}^{(t)\star} \right\rVert^2 \leqslant \frac{4\left\lVert \phi_{\text{C}}^{(0)} - \phi^{\star}_{\text{C}\vert\mu^{(0)}} \right\rVert^2}{\alpha_{1}\kappa_{\text{C}}} \nonumber\\
&+ \left[ \frac{2\alpha_{2}\kappa_{\text{C}}^{2}}{N_{\text{b}}} + \frac{2\alpha_{3}\kappa_{\text{A}}^{2}}{\kappa_{\text{C}}} \left( \alpha_{4} \kappa^2 + \frac{\alpha_{5}}{N_{\text{b}}} \right) \right] \frac{4T}{\alpha_{1}\kappa_{\text{C}}}\nonumber\\
&+ \frac{\alpha_{6}\kappa_{\text{A}}^{2}}{\kappa_{\text{C}}^{2}} \sum_{t = 0}^{T - 1} \mathbb{E} \left\lVert \nabla J(\mu^{(t)}) \right\rVert^2,
\end{align}
which is related to the convergence speed of the actor's update towards the stationary point, with $\alpha_{6}$ being another constant.

Ultimately we establish the overall convergence to a stationary policy by mitigating the cumulative tracking error through the actors' update process. By examining the relationship between the progress of the loss function and the tracking error, we can bound the cumulative policy gradient by
\vspace{-0.2cm}
    \begin{align}
    \frac{\kappa_{\text{A}}}{4} \sum_{t = 0}^{T - 1} &\mathbb{E} \left\lVert \nabla J(\mu^{(t)}) \right\rVert^2 \leqslant \frac{3\alpha_{4}\kappa_{\text{A}}}{4} \sum_{t = 0}^{T-1} \mathbb{E} \left\lVert \phi_{\text{C}}^{(t)}- \phi_{\text{C}\vert \mu}^{(t)\star} \right\rVert^2  \nonumber\\
    &+ \frac{R_{\text{max}}}{1 \!-\! \gamma} \!+ \!\frac{3 \kappa_{\text{A}}}{4} \left( \alpha_{4} \kappa^2 + \frac{\alpha_{5}}{N_{\text{b}}} \right) T.
\end{align}
Utilizing the relationship between the cumulative tracking error and the cumulative policy gradient, the tracking error can be ultimately canceled by the actor's positive progress towards a stationary point:
\vspace{-0.2cm}
\begin{align}
    &\left( \frac{\kappa_{\text{A}}}{4} - \frac{3\alpha_{6}\kappa_{\text{A}}^{3}}{4\kappa_{\text{C}}^{2}} \right) \sum_{t = 0}^{T - 1} \mathbb{E} \left\lVert \nabla J(\mu^{(t)}) \right\rVert^2 \nonumber\\
    &\leqslant \frac{R_{\text{max}}}{1 - \gamma} + \frac{3\alpha_{4}\kappa_{\text{A}}}{\alpha_{1}\kappa_{\text{C}}} \left\lVert \phi_{\text{C}}^{(0)} - \phi^{\star}_{\text{C}\vert\mu^{(0)}} \right\rVert^2 \nonumber\\
    &+ \left[ \frac{2 \alpha_2 \kappa_{\text{C}}^{2}}{N_{\text{b}}} + \frac{2\alpha_{3}\kappa_{\text{A}}^{2}}{\kappa_{\text{C}}} \left( \alpha_{4} \kappa^2 + \frac{\alpha_{5}}{N_{\text{b}}} \right) \right]\frac{3\alpha_{4}\kappa_{\text{A}}}{\alpha_{1}\kappa_{\text{C}}}T \nonumber\\
    &+ \frac{3 \kappa_{\text{A}}}{4} \left( \alpha_{4} \kappa^2 + \frac{\alpha_{5}}{N_{\text{b}}} \right)T.
\end{align}
Finally, by rearranging these terms, Eq.~(\ref{equ:pro_2}) in Proposition 2 can be obtained.




\ifCLASSOPTIONcaptionsoff
  \newpage
\fi

\vspace{-0.1cm}
\bibliographystyle{IEEEtran}
\bibliography{IEEEabrv,myrefs}

\end{document}